\documentclass[namedreferences]{solarphysics}
\usepackage[hyperref, optionalrh, natbib]{spr-sola-addons} 
\usepackage{graphicx}                    
\usepackage{color}   
\usepackage{breakurl}                    

\ifx \urlurl  \undefined \def \urlurl#1{\href{http://#1}{\textsf{#1}}}\fi

\newcommand{\degree}{^{\circ}}

\newcommand{\Rsun}{R$_{\odot}$}
\newcommand{\eg}{\textit{e.g.}}
\newcommand{\ie}{\textit{i.e.~}}

\newcommand{\vs}{\textit{vs.~}}

\newcommand{\cw}[1]{{\color{white}{#1}}}
\definecolor{light-gray}{gray}{0.30}
\begin{document}

\begin{article}

\begin{opening}

\title{First Simultaneous Views of the Axial and Lateral Perspectives of a Coronal Mass Ejection}


\author[addressref={aff1,aff2},corref,email={icabello@mendoza-conicet.gob.ar}]{\inits{I.}\fnm{I.}~\lnm{Cabello}}
\author[addressref={aff1,aff2},email={hebe.cremades@frm.utn.edu.ar}]{\inits{H.}\fnm{H.}~\lnm{Cremades}}
\author[addressref={aff3,aff4},email={lbalmaceda@icate-conicet.gob.ar}]{\inits{L.}\fnm{L.}~\lnm{Balmaceda}}
\author[addressref={aff1},email={ivodohmen@gmail.com}]{\inits{I.}\fnm{I.}~\lnm{Dohmen}}

\address[id=aff1]{Universidad Tecnol\'ogica Nacional - Facultad Regional Mendoza, Mendoza, Argentina}
\address[id=aff2]{Consejo Nacional de Investigaciones Cient\'ificas y T\'ecnicas (CONICET), Argentina}
\address[id=aff3]{Instituto Nacional de Pesquisas Espaciais (INPE), S\~ao Jos\'e dos Campos, Brazil}
\address[id=aff4]{Instituto de Ciencias Astron\'omicas, de la Tierra y del Espacio (ICATE) - CONICET, San Juan, Argentina}

%
\runningauthor{Cabello et al.}
\runningtitle{Simultaneous Views of Both Perspectives of a CME}

\begin{abstract}

The different appearances exhibited by coronal mass ejections (CMEs) are believed to be in part the result of different orientations of their main axis of symmetry, consistent with a flux-rope configuration. There are observational reports of CMEs seen along their main axis (axial perspective) and perpendicular to it (lateral perspective), but no simultaneous observations of both perspectives from the same CME have been reported to date. The stereoscopic views of the telescopes onboard the {\it Solar-Terrestrial Relations Observatory} (STEREO) twin spacecraft, in combination with the views from the {\it Solar and Heliospheric Observatory} (SOHO) and the {\it Solar Dynamics Observatory} (SDO), allow us to study the axial and lateral perspectives of a CME simultaneously for the first time. In addition, this study shows that the lateral angular extent (\emph{L}) increases linearly with time, while the angular extent of the axial perspective (\emph{D}) presents this behavior only from the low corona to $\approx$\,5 $R_{\odot}$, where it slows down. The ratio $L/D \approx$\,1.6 obtained here as the average over several points in time is consistent with measurements of \emph{L} and \emph{D} previously performed on events exhibiting only one of the perspectives from the single vantage point provided by SOHO.

\end{abstract}
%
\keywords{Coronal Mass Ejections, Initiation and Propagation; Coronal Mass Ejections, Low Coronal Signatures; Prominences, Dynamics}
\end{opening}


%
\section{Introduction}
\label{S_intro} 

Coronal mass ejections (CMEs) have been intensively studied since their first detections by space-borne coronagraphs, partly because they constitute the main modifier of heliospheric conditions, and hence of space weather. Given the adverse consequences that geomagnetic storms may unleash on Earth \citep[\eg][]{Lanzerotti2009}, the field of space weather has thrived. 
Unfortunately, to date it is not possible to predict when and where in the Sun the next eruption will take place. Therefore, current forecasting commences when a CME event has already been launched with a significant propagation component in Earth's direction. Throughout the years, numerous efforts have been undertaken to forecast the time of arrival of an interplanetary CME, the probability of it interacting with the Earth's magnetosphere, and the strength of this interaction \citep[\eg][]{Dryer-Smart1984, Fry-etal2001, Gopalswamy-etal2000, Gopalswamy-etal2001b, Smith-etal2003, Schwenn-etal2005, Gopalswamy-etal2005b, Gopalswamy-etal2005c, Manoharan2006, Taktakishvili-etal2009, Kilpua-etal2009, Moestl-etal2013, Xie-etal2013, Lugaz-etal2014}. As part of these efforts, it is crucial to understand how magnetic fields are organized within CMEs, and how this arrangement relates to the CME sources on the Sun. In this respect, it is fundamental to gain understanding of the general morphology of CMEs, as well as of how this morphology evolves with time.

Until the advent of the STEREO Mission at the end of 2006 \citep[{\it Solar-Terrestrial Relations Observatory};][]{Kaiser-etal2008}, the study of the three-dimensional (3D) configuration of CMEs had been speculative to some extent, given the limitations imposed by perspective and projection effects, inherent to bidimensional images obtained from a single vantage point. Some studies dealt with the analysis of observational properties of CMEs to deduce whether they were planar or 3D entities, and in the latter case evaluating whether the 3D overall structure was better approximated by spherically symmetric bubbles, by cylindrically-symmetric arcades, or by curved flux-tubes \citep[\eg][]{Crifo-etal1983, Schwenn1986, Webb1988, MacQueen1993, Vourlidas-etal2000, Plunkett-etal2000, Moran-Davila2004}. On the theoretical side, much progress has been made on 3D magnetohydrodynamic models that describe the initiation, eruption, configuration, and/or evolution of CMEs \citep[\eg][]{Gibson-Low1998, Antiochos-etal1999, Amari-etal2000, Tokman-Bellan2002, Manchester-etal2004, Torok-Kliem2005, Odstrcil-etal2005, Amari-etal2007, Zuccarello-etal2009}. Other models of geometrical basis have also proliferated \citep[\eg][]{Zhao-etal2002, Michalek-etal2003, Xie-etal2004, Cremades-Bothmer2005, Thernisien-etal2006}. The new views of the solar corona from different points of view provided after STEREO's launch certainly meant a step forward toward determining the 3D spatial extent of a CME and its true propagation direction \citep[\eg][]{Webb-etal2009, Mierla-etal2010, Mierla-etal2011, Gopalswamy-etal2012, Feng-etal2013}. The simultaneous two perspectives of the STEREO spacecraft also enabled the development of forward-modeling techniques \citep{Thernisien-etal2009, Wood-etal2010} that match well the appearance exhibited by a CME from both STEREO viewpoints and in some cases from the Earth's perspective as well. However, careless use of these tools may yield misleading reconstructions, given that at times it is possible to match several combinations of parameters to the same CME observations. As pointed out by \citet{Mierla-etal2009}, the 3D reconstruction of the CME morphology from currently available coronagraph data is an intrinsically undetermined task, given that a proper tomographic reconstruction requires a large number of images of a CME from many different viewpoints. 

\begin{figure}
 \centering
\includegraphics[width=0.55\textwidth]{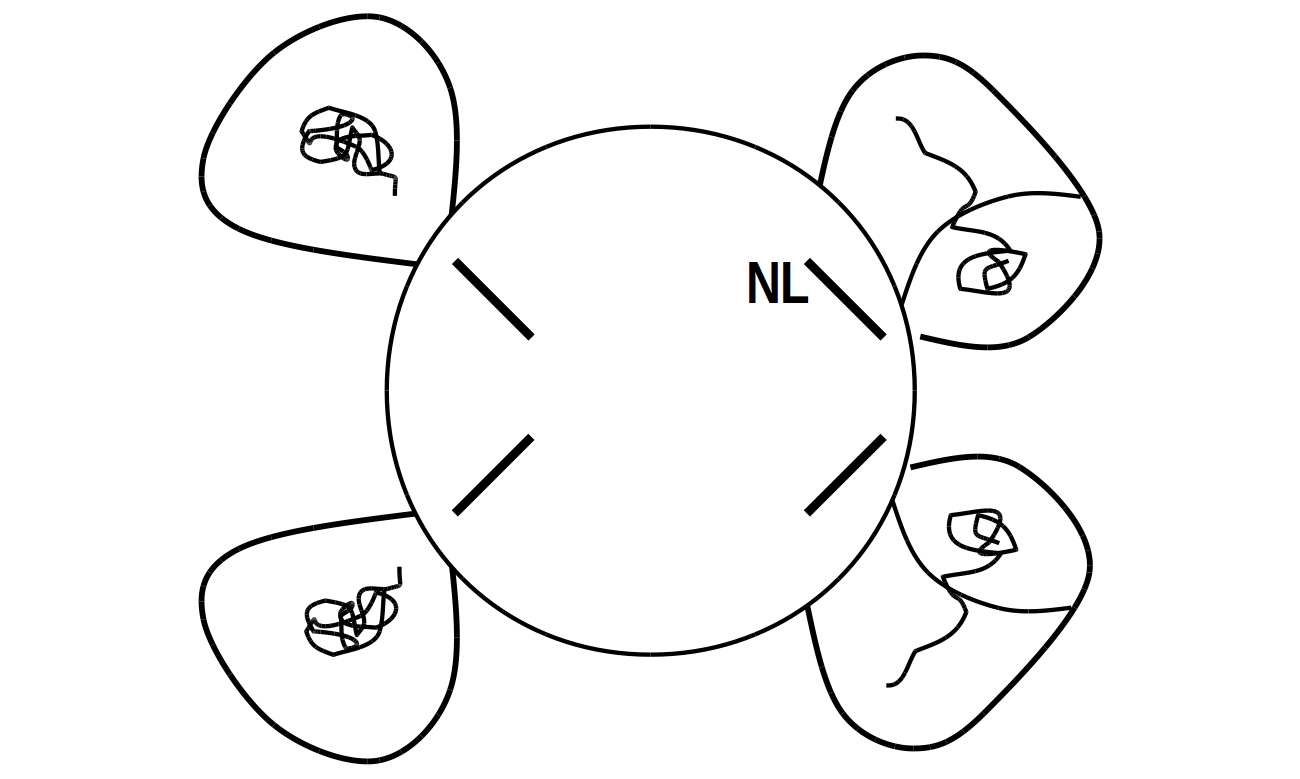}\llap{\parbox[b]{2.265in}{\tiny{\fontfamily{cmss}\selectfont Axial}\\\rule{0ex}{0.76in}}}\llap{\parbox[b]{2.50in}{\tiny{\fontfamily{cmss}\selectfont Perspective}\\\rule{0ex}{0.65in}}}\llap{\parbox[b]{0.50in}{\color{black}{\tiny{\fontfamily{cmss}\selectfont Lateral}}\\\rule{0ex}{0.76in}}}\llap{\parbox[b]{0.50in}{\color{black}{\tiny{\fontfamily{cmss}\selectfont Perspective}}\\\rule{0ex}{0.65in}}}

 \caption{The scheme of 3D configuration, adapted from \citet{Cremades-Bothmer2004}. NL stands for neutral line.}
 \label{scheme}
\end{figure}

In this effort we take advantage of coronal images provided by the STEREO spacecraft in  quadrature to investigate in detail the dimensions of a particular CME. The analysis relies on the overall 3D configuration scheme of cylindrical symmetry proposed by \citet{Cremades-Bothmer2004}, which approximates the structure of CMEs as organized along a main axis of symmetry, in agreement with the flux rope concept as described by \eg~\citet{Gosling-etal1995}, \citet{Chen-etal1997}, and \citet{Dere-etal1999}. The scheme considers the white-light topology of a CME projected in the plane of the sky (POS) as being primarily dependent on the orientation and position of the source region's neutral line on the solar disk. As a result of the solar differential rotation, the neutral lines associated with bipolar regions that are on the visible side of the solar disk and close to the east limb tend to be perpendicular to the limb, while when close to the west limb, the neutral lines tend to be parallel to it (see Figure~\ref{scheme}). According to these typical orientations, front-side solar sources close to the east limb tend to yield CMEs seen along their main axis, exhibiting a three-part structure and in many cases also helical threads indicative of magnetic flux ropes as in \eg~\citet{Wood-etal1999} and \citet{Dere-etal1999}. On the other hand, front-side solar sources close to the west limb tend to yield CMEs with their main axes oriented parallel to the limb and perpendicular to the observer--Sun line, so that the lateral view of a CME is detected. In view of this cylindrically symmetric configuration, \citet{Cremades-Bothmer2005} measured the lateral angular extent \emph{L} of the cylinder axis and the angular extent of the cylinder cross section \emph{D} on SOHO/LASCO CMEs exhibiting extreme projections, \ie seen solely either in the lateral or in the axial perspective, respectively. These angular extents were dissimilar on average for both groups, \ie CMEs seen along their symmetry axis (axial events) appeared narrower than those seen perpendicular to it (lateral events). However, it remains unknown whether this trend is verified for all CMEs or if it was only fortuitous, given that the measurements of \emph{L} and \emph{D} have thus far not been performed simultaneously on the same CME. As argued in Section \ref{s:ID}, it is very difficult to detect a CME event that exhibits both perspectives simultaneously. To our knowledge, this is the first time that both \emph{L} and \emph{D} are reported to be simultaneously measured for the same event. The next section describes the investigated data sets, while Section \ref{s:ID} addresses the criteria considered to identify this singular event. Section \ref{s:3Dmodel} presents the modeling of this event using the GCS forward model \citep[Graduated Cylindrical Shell;][]{Thernisien-etal2009,Thernisien2011}. The characterization of the angular extents is presented in Section \ref{s:expansion}, while Section \ref{s:conclusions} presents final remarks and conclusions.

\section{Analyzed Data}\label{s:Data} 

The primary data used in the present work were supplied by SECCHI \citep[\emph{Sun-Earth Connection Coronal and Heliospheric Investigation};][]{Howard-etal2008} onboard the STEREO spacecraft, in particular by the COR1 and COR2 coronagraphs. The STEREO Mission consists of two identical spacecraft following Earth's orbit around the Sun, one ahead of it (ST-A) and the other behind it (ST-B). The spacecraft drift away from the Sun-Earth line at a rate of $\approx$\,22$\degree$ {\it per} year. Background images, obtained as indicated in \citet{Thompson-etal2010}, were used to remove most of the F-corona and stray light. Data from the C2 and C3 coronagraphs from the LASCO \citep[\emph{Large Angle Spectroscopic Coronagraph};][]{Brueckner-etal1995} experiment onboard SOHO \citep[\emph{Solar and Heliospheric Observatory};][]{Domingo-etal1995} were used as well.

To identify the source region of the event, data from the EUVI (\emph{Extreme-Ultraviolet Imager}) instrument onboard SECCHI at the 195\,\AA~ and 304\,\AA~ emission lines were inspected. Additionally, 193\,\AA~low-coronal images from the AIA \citep[\emph{Atmospheric Imaging Assembly};][]{Lemen-etal2012} instrument onboard the SDO \citep[\emph{Solar Dynamics Observatory};][]{Pesnell-etal2012} spacecraft were examined. H$\alpha$ data from the Paris-Meudon spectroheliograph at Pic Du Midi Observatory (http://bass2000.obspm.fr/), and from the \emph{New H$\alpha$ Patrol Telescope} at Big Bear Solar Observatory (BBSO; http://www.bbso.njit.\\
edu/) were also analyzed.

To highlight structures and track the evolution of features in a sequence of images, difference images were produced by subtracting consecutive images (running-difference) as well as by subtracting a pre-event image from the whole sequence (base-difference).

\section{Identification of the Event}\label{s:ID}
Coronagraphs provide views of CMEs projected in their POS, thus offering 2D images that hinder the study of the 3D configuration of CMEs.
As pointed out by \cite{Cremades-Bothmer2004,Cremades-Bothmer2005}, a CME can exhibit a very different  appearance depending on the point of view; the two archetypical perspectives are the axial and lateral ones. Stereoscopic images from the STEREO mission, in combination with the terrestrial views from SOHO and SDO, allow the simultaneous study of a CME from different vantage points. In spite of this, no simultaneous observations of the axial and lateral perspectives of a given event have been reported before. Such detections are rarely possible, given that it is required (i) that at least two spacecraft are approximately in quadrature, (ii) that the event propagates nearly perpendicular to the plane that contains the spacecraft, and (iii) that the main axis of the CME has a particular orientation with respect to the observers. The fact that CMEs can be ejected in any direction and from any position in the solar disk, with their axis orientation being also variable, hinders the simultaneous observation of these two archetypical perspectives. 

\begin{figure}
   \centering
\begin{tabular}{cc}
{\includegraphics[width=5cm]{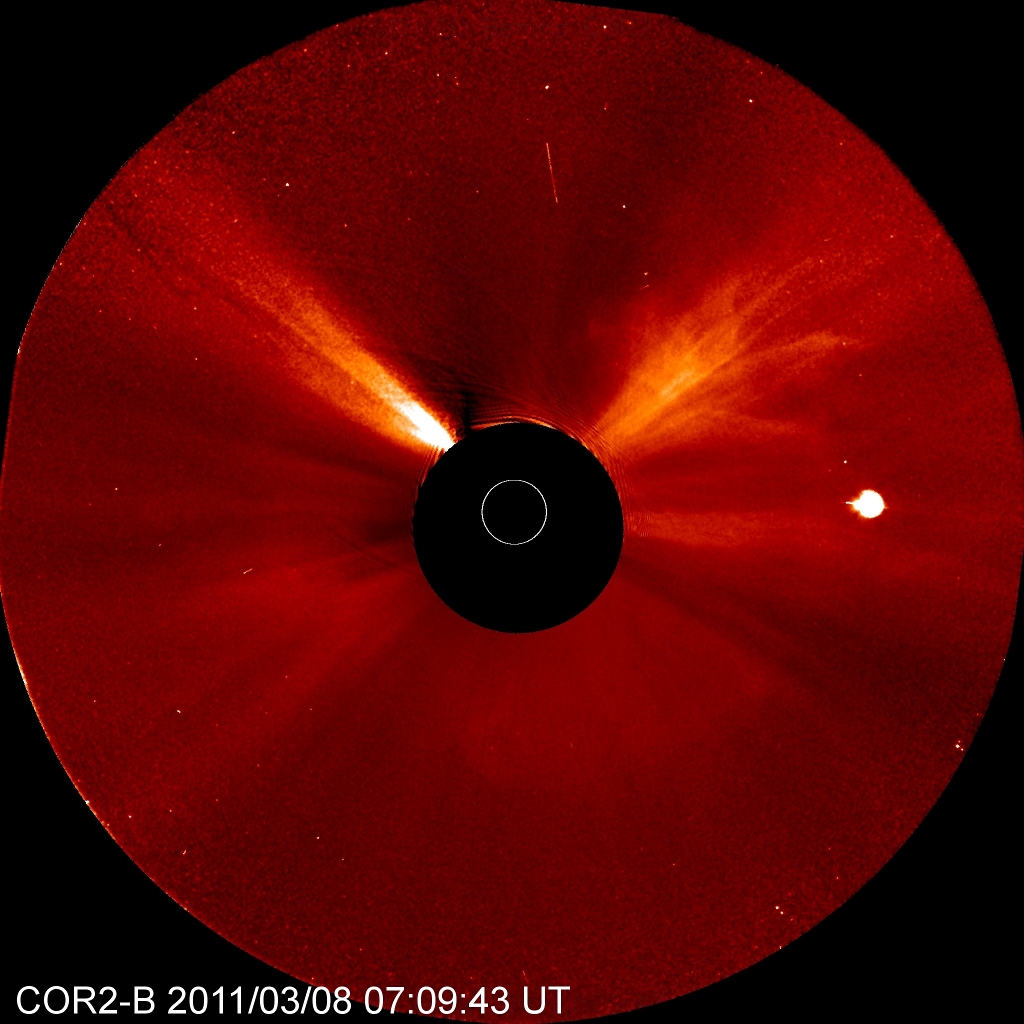}}
   & {\includegraphics[width=5cm]{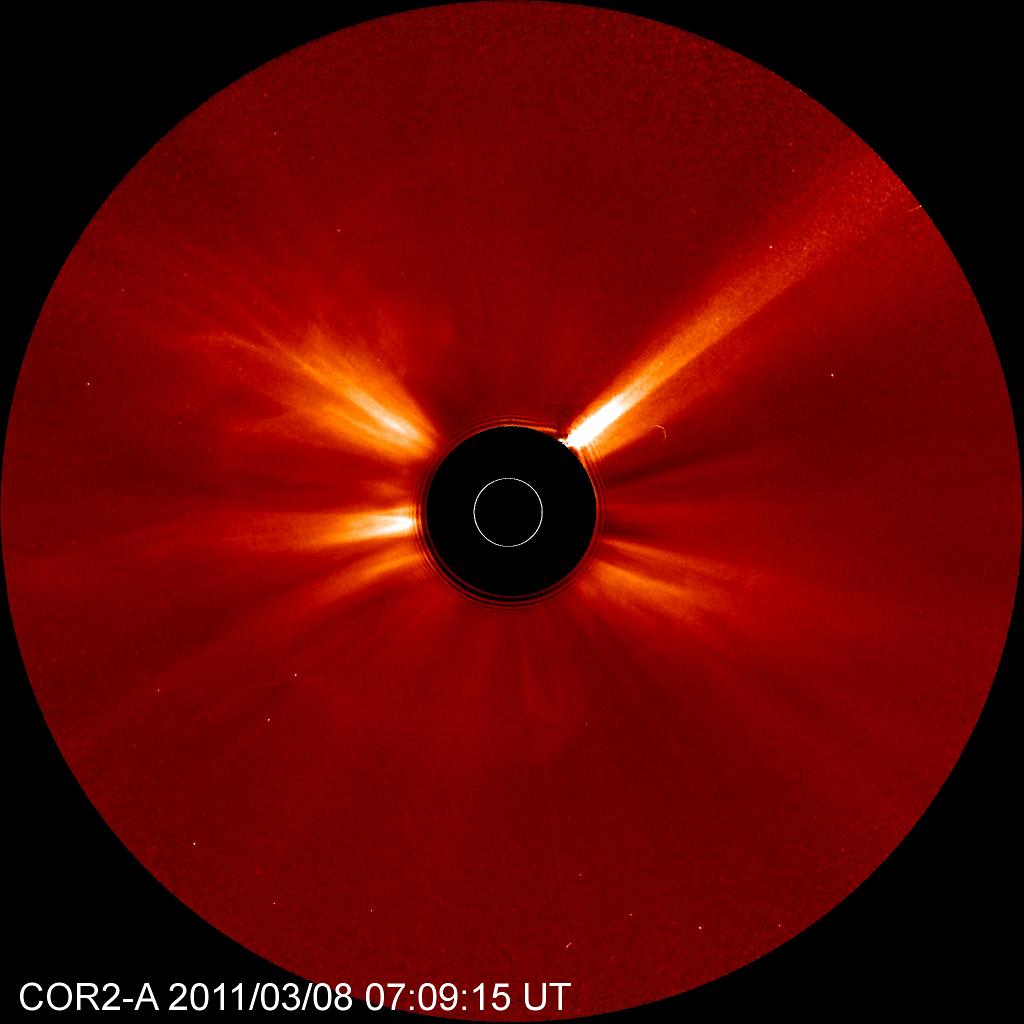}}\\

{\includegraphics[width=5cm]{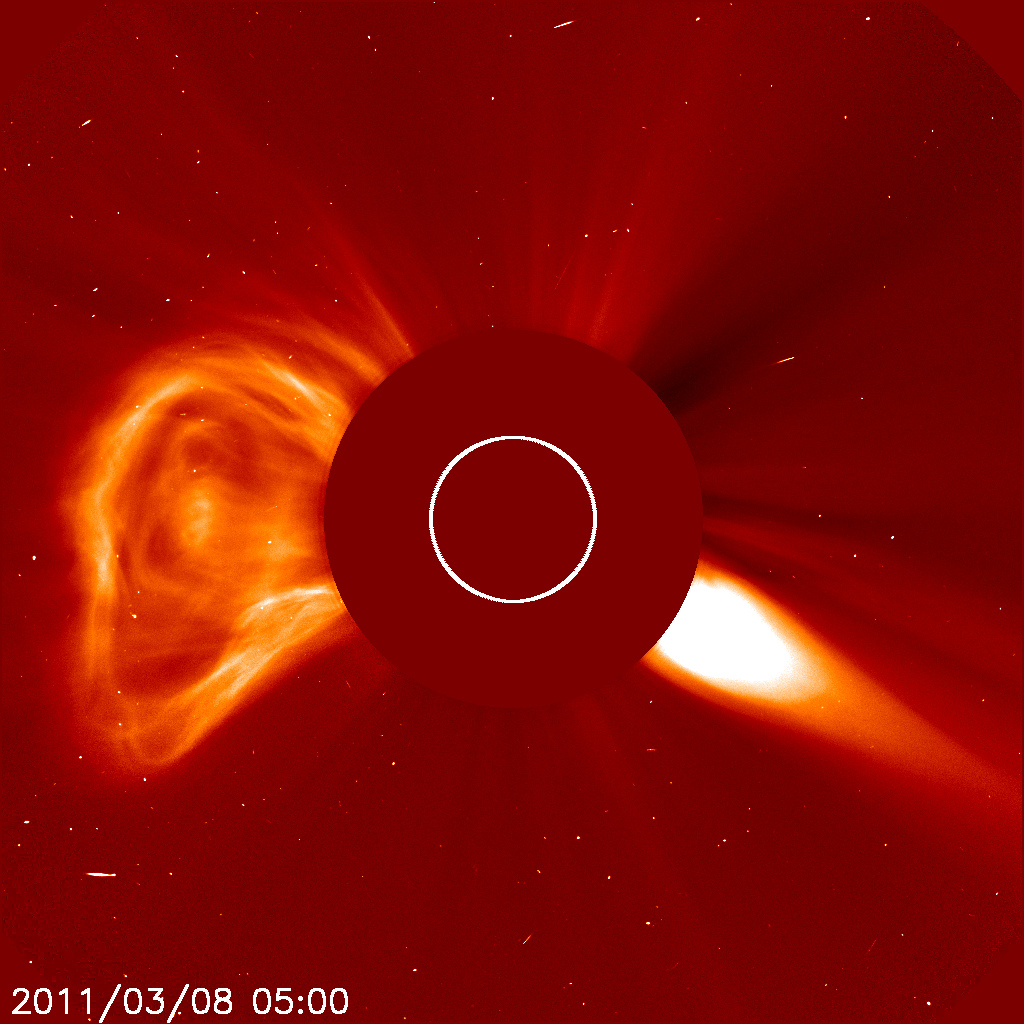}} 
   &{\includegraphics[width=5cm]{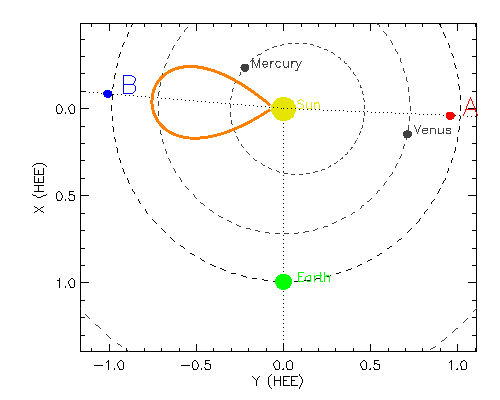}}\\
\end{tabular}

\caption{Three nearly simultaneous views of a CME aimed toward ST-B on 8 March 2011, with the STEREO spacecraft $\approx$\,180$\degree$ apart and $\approx$\,90$\degree$ away from Earth. From left to right and top to bottom: views from ST-B COR2, ST-A COR2, and SOHO/LASCO C2, and positions of Earth and the STEREO spacecraft in a top view of the ecliptic. The STEREO Orbit Tool is available at the STEREO Science Center (\url{http://stereo-ssc.nascom.nasa.gov}).}
\label{20110308}
\end{figure}

\begin{figure}
   \centering
\begin{tabular}{cc}
{\includegraphics[width=5cm]{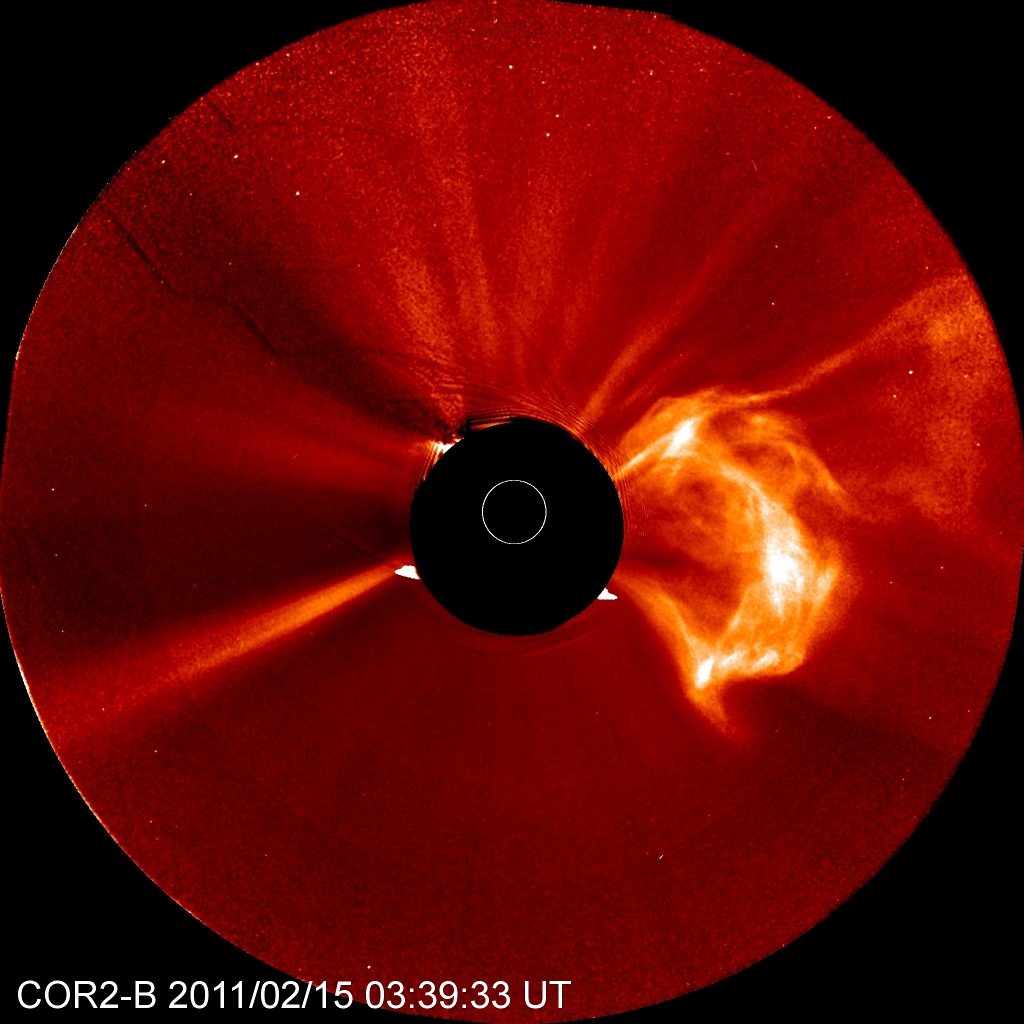}} 
   & {\includegraphics[width=5cm]{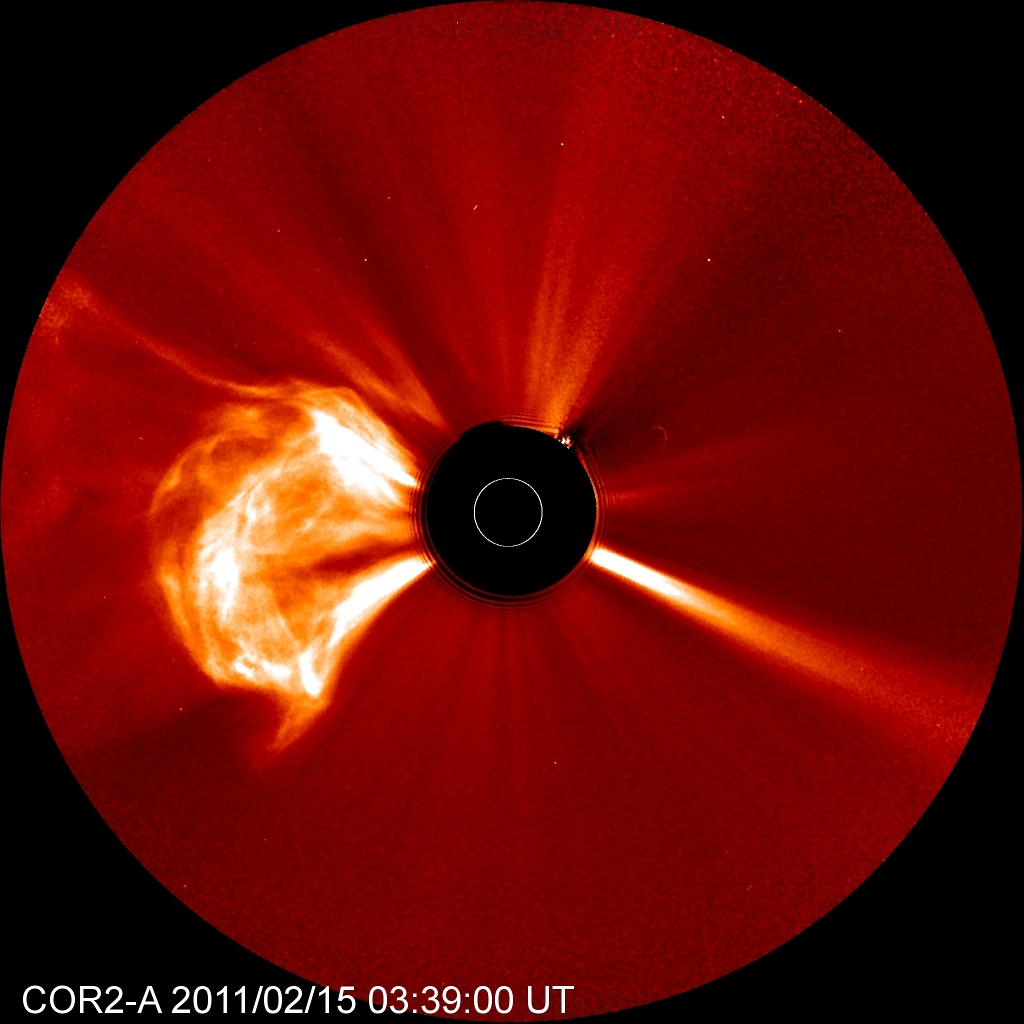}}\\
   
{\includegraphics[width=5cm]{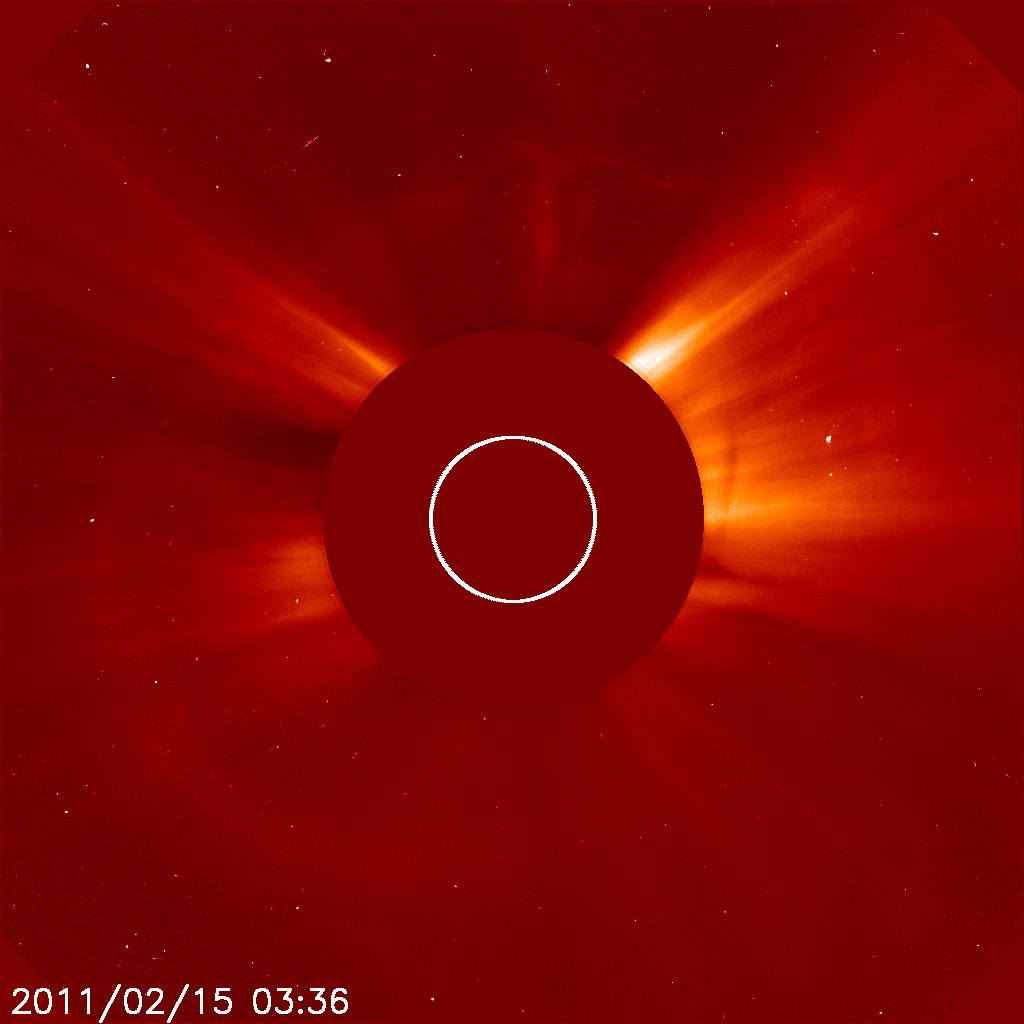}} 
   &{\includegraphics[width=5cm]{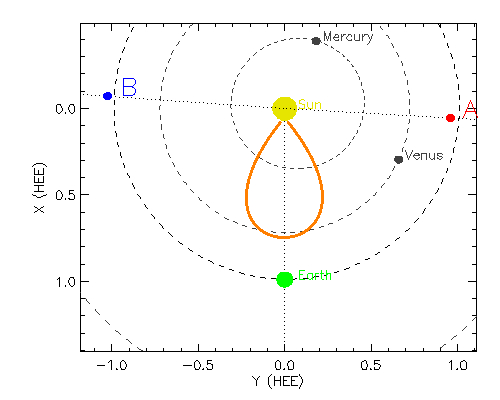}}\\
\end{tabular}

\caption{Same as Figure~\ref{20110308} but with the Earth-directed CME of 15 February 2011, under a similar spacecraft configuration.}
\label{20110215}
\end{figure}

\begin{figure}
   \centering
\begin{tabular}{cc}
{\includegraphics[width=5cm]{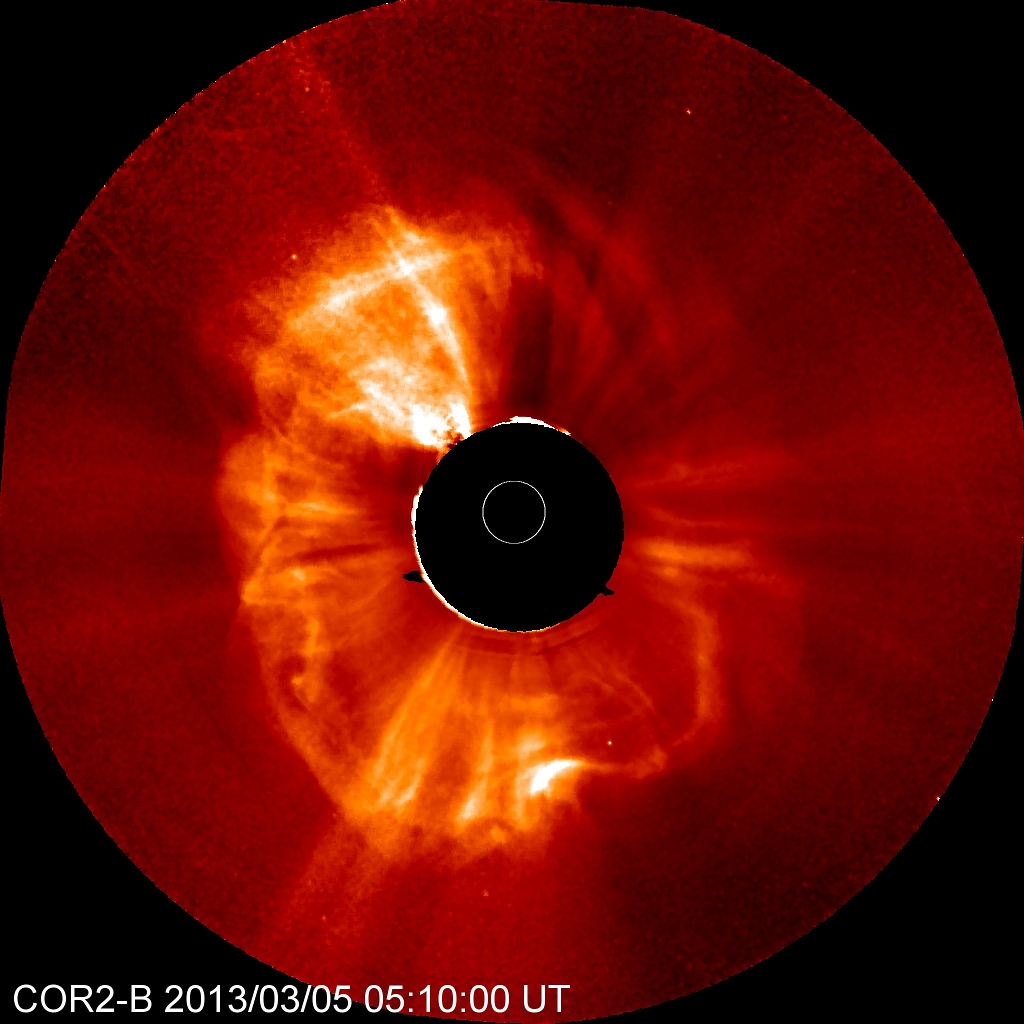}} 
   & {\includegraphics[width=5cm]{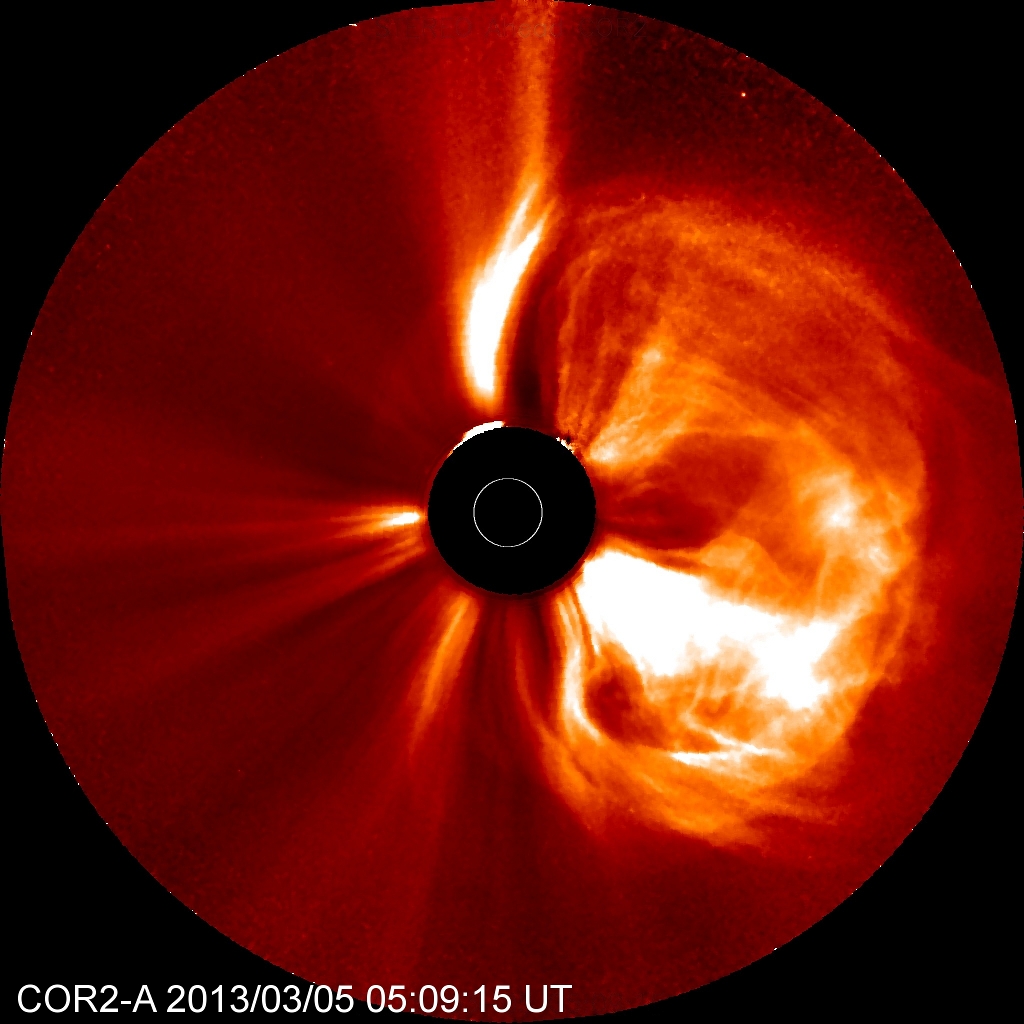}}\\
   
{\includegraphics[width=5cm]{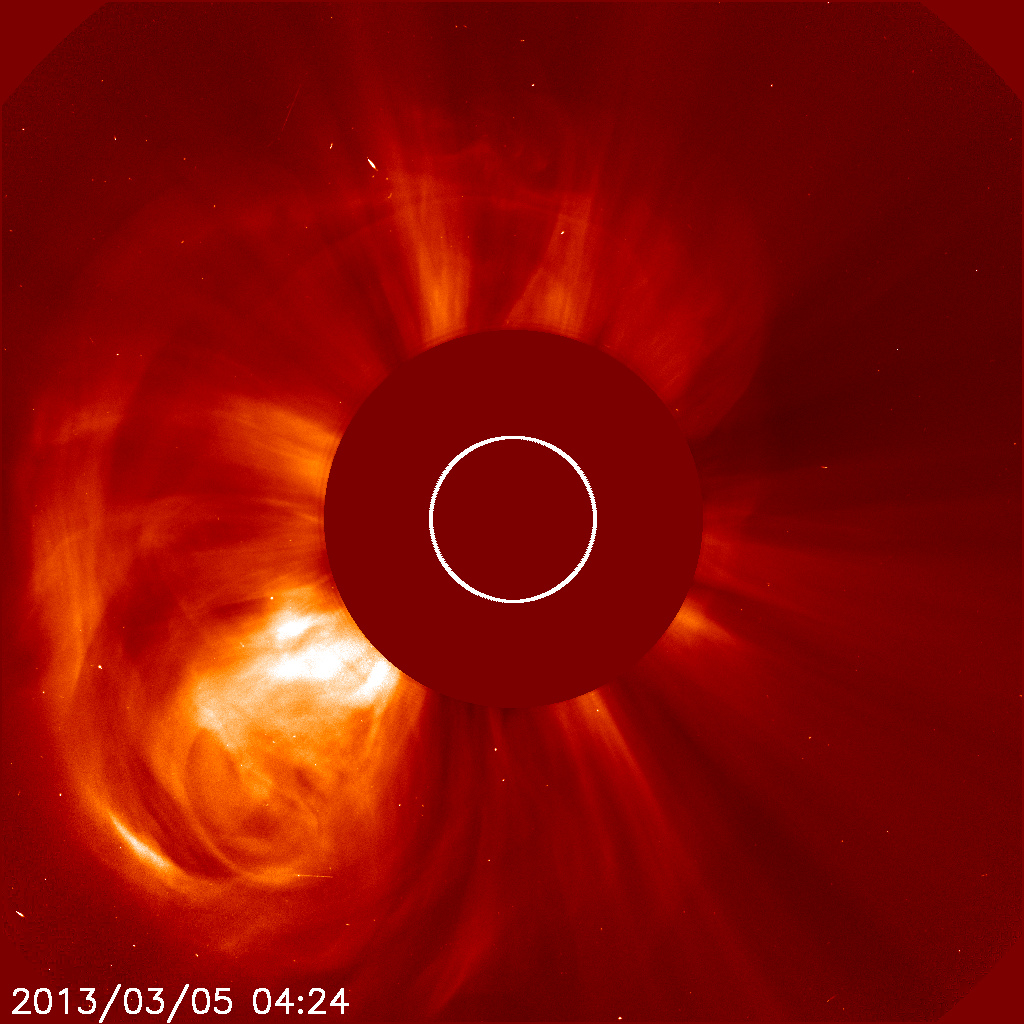}} 
   &{\includegraphics[width=5cm]{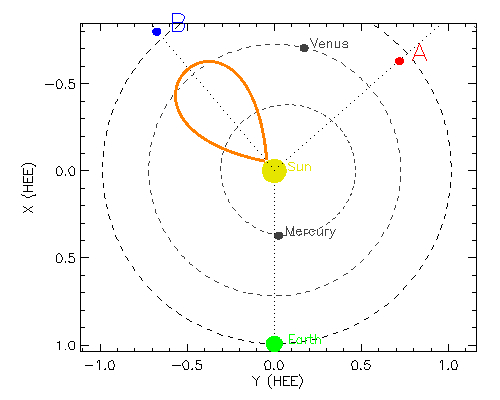}}\\
\end{tabular}

\caption{Same as Figure~\ref{20110308} but for the CME on 5 March 2013, with the STEREO spacecraft $\approx$\,90$\degree$ apart, and $\approx$\,135$\degree$ away from Earth.}
\label{20130305}
\end{figure}

Figures \ref{20110308}\,-\,\ref{20130305} are examples that show quadrature situations that allow the observation of only one perspective. In the three figures, the images correspond to: ST-B COR2 (top left), ST-A COR2 (top right), SOHO/LASCO C2 (bottom left), and the position of the STEREO and SOHO spacecraft (bottom right). In Figure~\ref{20110308}, the STEREO spacecraft are $\approx$\,180$\degree$ apart and $\approx$\,90$\degree$, \ie in quadrature, with respect to SOHO. A CME is directed toward ST-B, so that it is observed from both STEREO spacecraft as a halo CME, \ie with a bright rim surrounding the coronagraph occulter, and none of the archetypical perspectives are discernible. At the same time, the CME approximately travels in the POS of SOHO/LASCO C2, so that it is detected as a limb CME. Depending on the orientation of the CME main axis, this spacecraft configuration allows observing either an intermediate view somewhere between the axial and the lateral perspectives, or at best only one of the archetypical perspectives in the SOHO/LASCO C2 field of view (FOV). For the case presented in Figure~\ref{20110215} the three spacecraft are almost equally distributed as in Figure~\ref{20110308}, but with the CME directed toward SOHO. Therefore, this CME is observed as a halo by SOHO/LASCO C2 and it travels almost in the POS of both STEREO coronagraphs. If the CME is conveniently oriented, at most either the axial or the lateral perspective would be simultaneously observed from both STEREOs. Figure~\ref{20130305} shows other spacecraft--CME configuration that is unsuitable for simultaneously observing both perspectives: the STEREO spacecraft are $\approx$\,90$\degree$ apart and a CME is directed toward ST-B. With such a configuration, a halo CME is observed from ST-B, and at most one of the archetypical perspectives can be detected from ST-A if the CME main axis is well oriented, while SOHO/LASCO observes an intermediate perspective. In general, with any two of these spacecraft separated $\approx$\,90$\degree$ and a CME directed toward any of them, it is possible to detect only one of the archetypical perspectives, either the axial or the lateral one, plus a halo CME.


\begin{figure}
   \centering
\begin{tabular}{cc}
{\includegraphics[width=5cm]{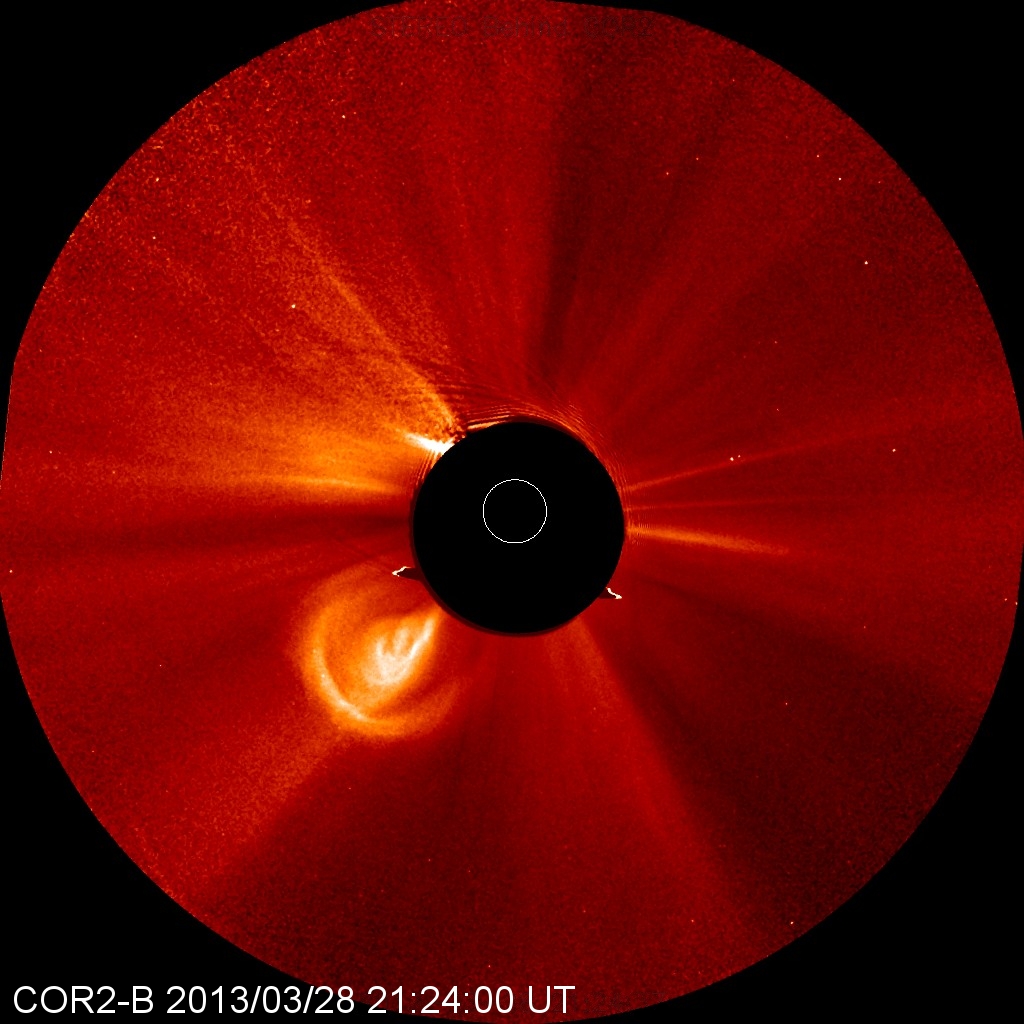}} 
   & {\includegraphics[width=5cm]{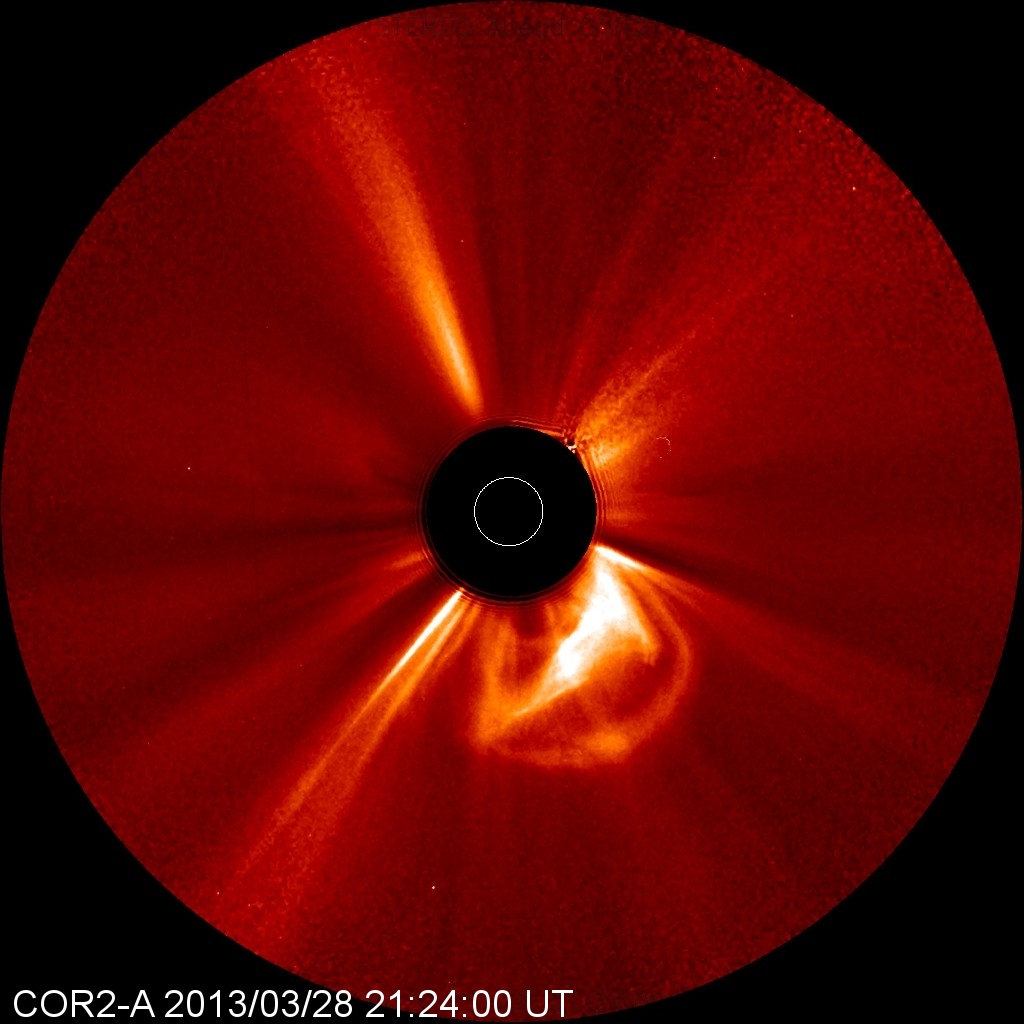}}\\
   
{\includegraphics[width=5cm]{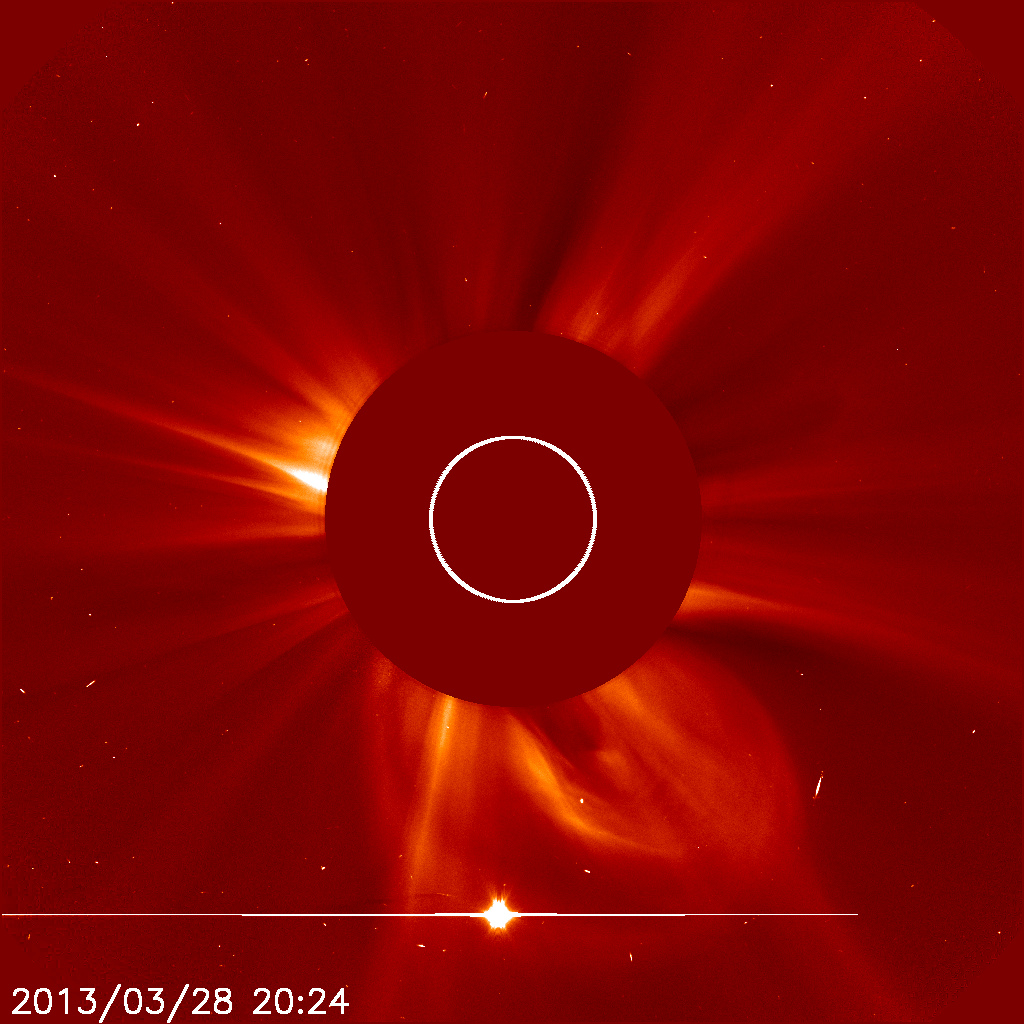}} 
   &{\includegraphics[width=5cm]{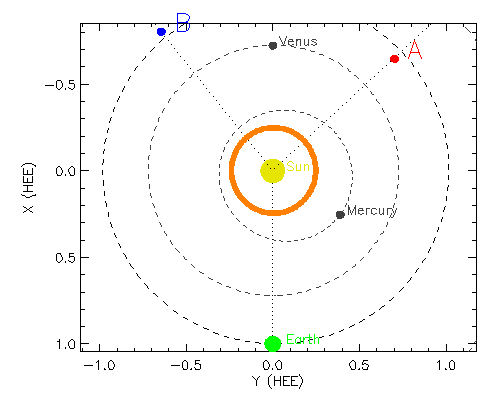}}\\
\end{tabular}

\caption{Same as Figure~\ref{20110308} but for the CME on 28 March 2013, with the STEREO spacecraft $\approx$\,90$\degree$ apart, and $\approx$\,135$\degree$ away from Earth.}
\label{20130328}
\end{figure}

Because of the observational limitations imposed by the orbits of these spacecraft, the only circumstances under which both perspectives can be simultaneously detected are given when a CME propagates nearly perpendicular to the ecliptic, with at least two spacecraft in quadrature. At the same time, the symmetry axis of the CME must be aligned with one of the the Sun--spacecraft lines, and perpendicular to the other. To identify such an event simultaneously showing both perspectives, the SOHO/LASCO CME Catalog \citep[\url{http://cdaw.gsfc.nasa.gov/CME_list};][]{Yashiro-etal2004} was inspected during time periods when two of the three spacecraft ST-A, ST-B, and SOHO were in quadrature, with either 180$\degree$$\pm$\,10$\degree$ (November 2010 - July 2011) or 90$\degree$$\pm$\,10$\degree$ (October 2008 - June 2009 and December 2012 - June 2013) between the first two. In addition, we considered events with a central position angle (PA) within $\pm$\,25$\degree$ with respect to the 0$\degree$ PA (north pole) and 180$\degree$ PA (south pole), as viewed from SOHO. A total of 38 events were identified that fulfilled these criteria. However, in this work we study in detail the south pole event on 28 March 2013 because the orientation of its main axis was highly favorable to be observed as axial by one spacecraft and lateral by the other. This event is first seen moving in the FOV of STEREO/SECCHI COR1 some time after 14:00 UT and appearing in SOHO/LASCO C2 around 17:00 UT. According to the CDAW SOHO/LASCO CME Catalog, its projected speed on the POS shows a second-order evolution, reaching a speed of 655~km~s$^{-1}$ at 20 \Rsun, with an acceleration of 15.9 m s$^{-2}$. Figure~\ref{20130328} shows the event as detected from the three spacecraft, and their spatial configuration with ST-B and ST-A $\approx$\,86$\degree$ apart. The bright core material is seen concentrated along the line of sight in the COR2-B image, given that the CME axis is aligned with the Sun--observer line. On the other hand, an extended core nearly perpendicular to the Sun--observer line, distinctive of the lateral perspective, is seen in the COR2-A image. An intermediate perspective is detected by SOHO/LASCO C2, given that the CME main symmetry axis is oriented $\approx$\,45$\degree$ from the Sun-Earth line. The three-part structure is discernible in the three views, although with different appearances. The bright core is frequently associated with erupting prominences, as reported by various authors \citep[\eg][]{Illing-Hundhausen1985,Vourlidas-etal2013,Webb2015}. 

\begin{figure}[!ht]
   \centering
   \includegraphics[trim = 75mm 0mm 75mm 250mm, clip,width=1\textwidth]{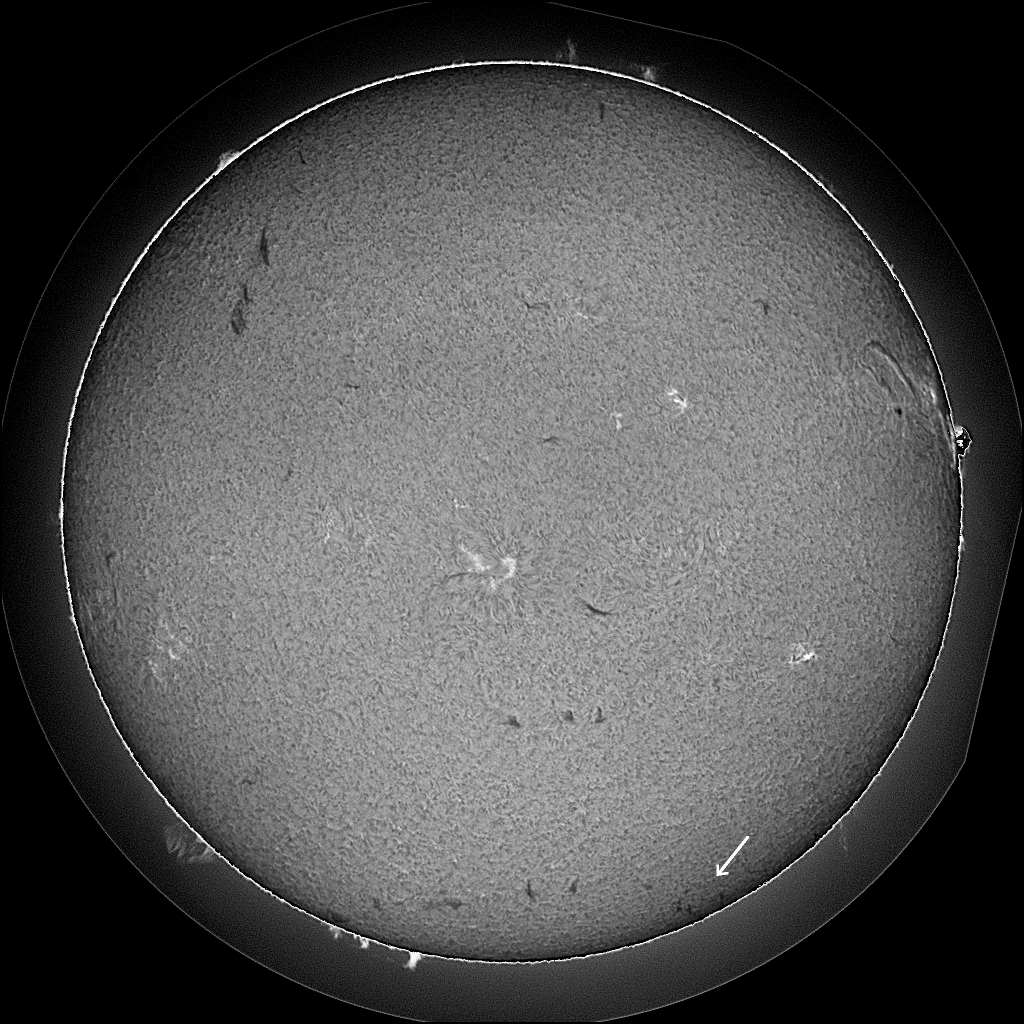}

   \caption{BBSO H$\alpha$ image recorded on 21 March 2013 at 22:48:54 UT showing the south polar crown filaments. Those at the southwest limb are presumably related to the event under study.}
   \label{halpha}
 \end{figure}

To identify the source region, we looked for evidence of eruptions in H$\alpha$ and EUVI images from SDO/AIA and STEREO/SECCHI. As a first step, H$\alpha$ data from the Paris-Meudon spectroheliograph and from the \emph{New H$\alpha$ Patrol Telescope} at BBSO were carefully scanned at latitudes and longitudes consistent with those of the event observable in LASCO C2 images. No evidence of filament disappearance or two-ribbon flare were found in H$\alpha$ images, in agreement with the fact that the source region was located behind the south pole limb as seen from Earth. From the COR2 images displayed in Figure~\ref{20130328} (top), a propagation direction with a small component away from Earth can be appreciated. In the figure, the CME is observed to propagate partially toward the east as seen from ST-B, and slightly toward the west from ST-A, both facts in agreement with a back-sided propagation as seen from Earth (\ie away from us). Considering solar rotation and spacecraft positions, H$\alpha$ images should show the potential source region at the southwest limb about a week earlier (see Figure~\ref{halpha}). It is highly likely that this CME is related to the polar crown filaments present at south polar latitudes. 

\begin{figure}[!h]
  \centering
  \includegraphics[trim = 15mm 65mm 0mm 0mm, clip,width=0.32\textwidth]{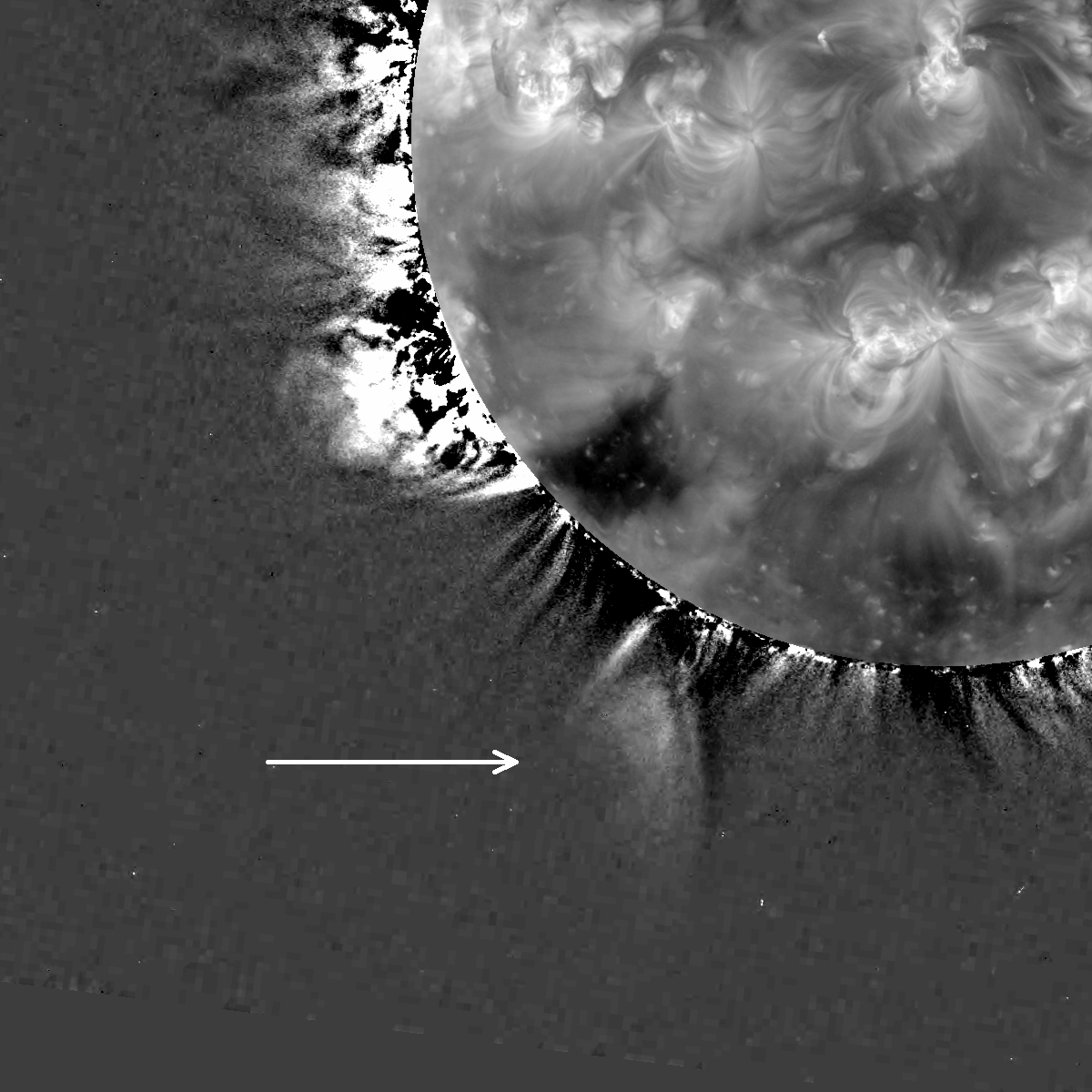}\llap{
  \parbox[b]{1.53in}{\colorbox{light-gray}{\color{white}{\tiny {\fontfamily{cmss}\selectfont EUVI-B 20130328 14:51 UT}}}\\\rule{0ex}{1.195in}}}
  \includegraphics[trim = 0mm 0mm 0mm 45mm, clip,width=0.32\textwidth]{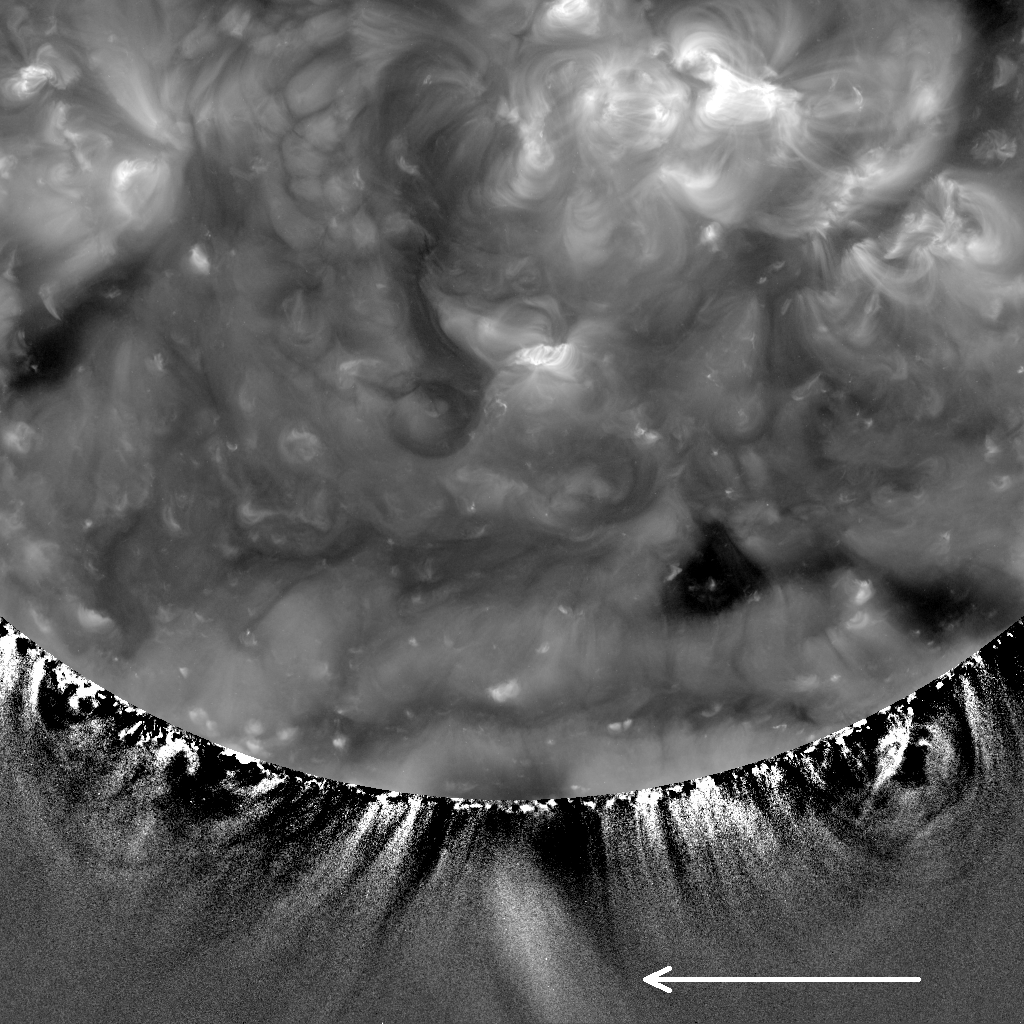}\llap{
  \parbox[b]{1.53in}{\colorbox{light-gray}{\color{white}{\tiny {\fontfamily{cmss}\selectfont AIA 20130328 14:52 UT}}}\\\rule{0ex}{1.195in}}}
  \includegraphics[trim = 0mm 0mm 0mm 52mm, clip,width=0.32\textwidth]{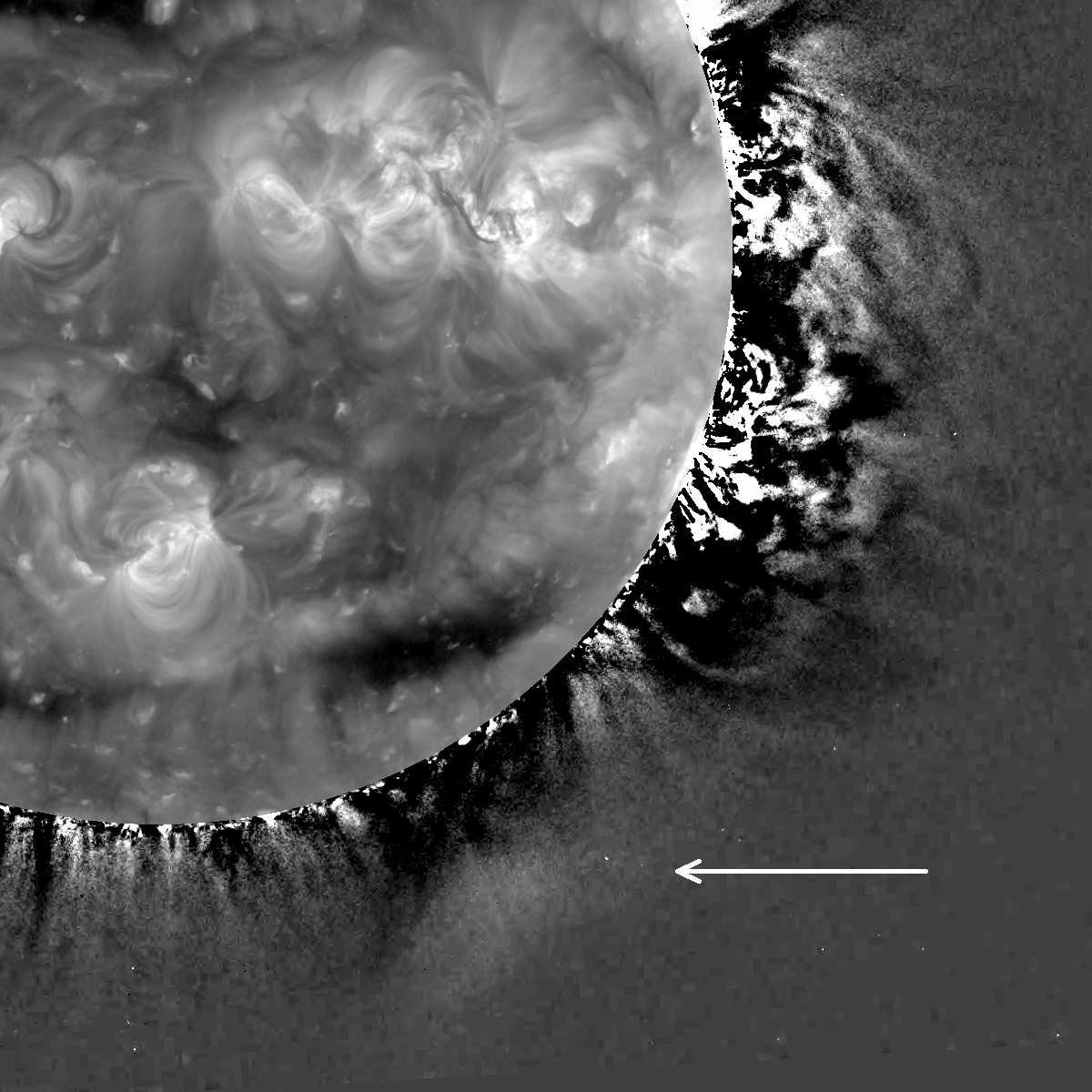}\llap{
  \parbox[b]{1.53in}{\colorbox{light-gray}{\color{white}{\tiny {\fontfamily{cmss}\selectfont EUVI-A 20130328 14:51 UT}}}\\\rule{0ex}{1.195in}}}
  \caption{Images captured on 28 March 2003 by STEREO/SECCHI EUVI 195\,\AA~ST-B (left) and ST-A (right), and by SDO/AIA 193\,\AA~(center). A dim feature indicated by arrows is seen in eruption by the three instruments. From 1\,\Rsun~onward, we show base-difference images to increase the contrast of the moving structure, while the solar disk is presented in direct images.}
  \label{low_corona}
\end{figure}

On the other hand, SDO/AIA 193\,\AA~as well as SECCHI/EUVI 195\,\AA~and 304\,\AA~images were carefully inspected, taking into account that a temporal and spatial correspondence exists between the EUV and the white-light features. No signs of eruptive or post-eruptive activity were detected on the surface in any of these. However, a dim feature was observed to slowly rise from the south limb in AIA 193\,\AA~images (see central panel in Figure~\ref{low_corona}). For completeness, we also examined images at 174\,\AA~provided by the SWAP instrument onboard the PROBA-2 mission \citep{Seaton-etal2013,Halain-etal2013}, but in a similar way, we found no traces of eruption on the disk, while we identified some outward-propagating material. As argued before, the fact that we cannot detect surface activity from the Earth's perspective is compatible with a back-sided propagation. From the vantage points of STEREO, the outward-propagating feature is observed to rise above the southeast and southwest limbs in difference images of EUVI-A and -B 195\,\AA, respectively (see left and right panels in Figure~\ref{low_corona}), without noticeable surface activity. This feature is seen to be stable for hours in EUVI-B 195\,\AA~images before the eruption. In EUVI 304\,\AA~images, only a small fraction of the prominence is seen to rise and disappear in the background. We argue that a polar crown filament potentially associated with this CME has remained in a quiescent state and at a high altitude for a long period of time, thus allowing for the filament to be heated to coronal temperatures and therefore turning it undetectable in the 304\,\AA~passband, developing into a sort of stealth CME \citep{Robbrecht-etal2009}. This hypothesis is supported by the dark feature observed in base-difference COR1 images, addressed in Section \ref{s:expansion}, which is indicative of a pre-existing structure at a relatively high altitude that was blown away during the eruption.

\section{3D Reconstruction}\label{s:3Dmodel}
\cite{Thernisien-etal2009} developed a forward-modeling tool to reproduce the three-dimensional configuration of CMEs using white-light data from the STEREO Mission. This method derives from the GCS model \citep{Thernisien-etal2006}, based on the findings by \cite{Cremades-Bothmer2004}. The forward-modeling tool reproduces the 3D flux-rope structure of a CME by considering it as a spring-shaped twisted flux tube. It is mainly designed as a horizontal tubular structure organized around a main axis, held on two cones that account for the legs, in a way that the ensemble has the shape of a hollow croissant. In spite of the success of this method to approximate the 3D flux-rope structure of CMEs, there are two main caveats to bear in mind: i) the internal structure of CMEs is not considered and ii) misusage of the tool is very frequent because its application is often taken lightly, \ie it may appear to the untrained eye that several different solutions fit the same CME.

\begin{figure}[!ht]
  \centering
  \includegraphics[width=1\textwidth]{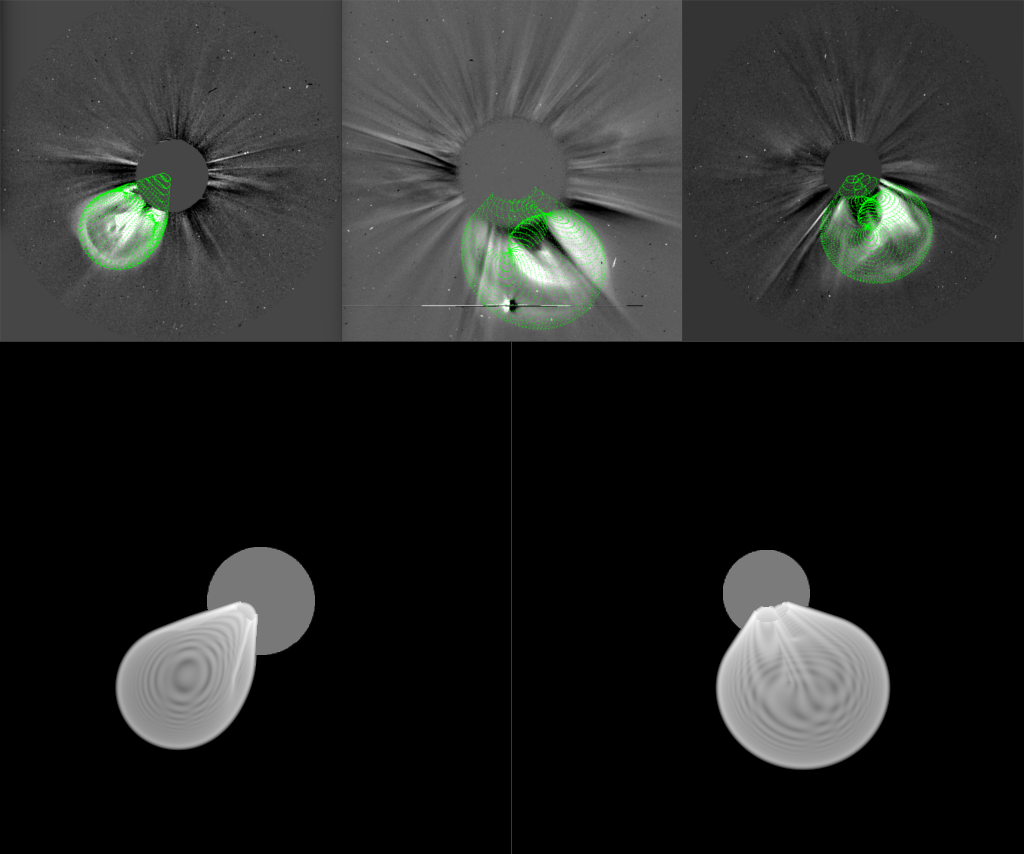}\llap{\parbox[b]{4.7in}{\tiny\cw{{\fontfamily{cmss}\selectfont COR2-B 20130328 22:24 UT}}\\\rule{0ex}{3.8in}}}\llap{\parbox[b]{3.1in}{\tiny \cw{{\fontfamily{cmss}\selectfont LASCO C2 20130328 20:24 UT}}\\\rule{0ex}{3.8in}}}\llap{\parbox[b]{1.5in}{\tiny \cw{{\fontfamily{cmss}\selectfont COR2-A 20130328 22:24 UT}}\\\rule{0ex}{3.8in}}}\llap{\parbox[b]{4.7in}{\tiny \cw{{\fontfamily{cmss}\selectfont COR2-B 20130328 22:24 UT (simulation)}}\\\rule{0ex}{2.15in}}}\llap{\parbox[b]{2.3in}{\tiny \cw{{\fontfamily{cmss}\selectfont COR2-A 20130328 22:24 UT (simulation)}}\\\rule{0ex}{2.15in}}}
  \caption{Top: Simulations of the CME outer envelope, performed with the forward-modeling tool based on the GCS model, are shown as a green mesh superimposed on the COR2-B (left), COR2-A (right), and LASCO C2 (center) base-difference images. Bottom: Synthetic white-light images from the perspective of the ST-B (left) and ST-A (right) spacecraft, with a gray circle representing the occulter area.}
  \label{20130328_synthetic}
\end{figure}

This forward-modeling tool has been used to reproduce the three observed perspectives of the CME under study, by adjusting six free parameters such that the GCS modeled envelope simultaneously fits the projected CME shape in the COR2-B, LASCO, and COR2-A images (see upper panels in Figure~\ref{20130328_synthetic}). The first set of parameters was obtained by fitting nearly simultaneous images at 22:24 UT from COR2-A, COR2-B, and LASCO C3. These are the height of the leading edge of the model $h_{front}$ = 12\,\Rsun; the half-angle between the cones $\alpha = 16\,^{\circ}$; the aspect ratio \mbox{$\kappa$ = a($r$) / $r$} = 0.47, where a($r$) is the varying radius of the cross section of the envelope at the distance $r$ from the Sun's center; the tilt angle around the axis of symmetry of the model $\gamma = 15\,^{\circ}$; the source region longitude $\phi = 158\,^{\circ}$ and the latitude $\theta = -44\,^{\circ}$, with the last two given in the Stonyhurst coordinate system \citep{Thompson2006}. The aspect ratio $\kappa$ is additionally related to the half-angle of each cone $\delta$ by $\delta = arcsin(\kappa)$, from where the angular width of the axial extent can be determined as $2\,\delta$, and that of the lateral extent as $2\,(\alpha\,+\,\delta)$ \citep{Thernisien2011}. The angular widths of both perspectives are addressed in Section \ref{s:expansion}.

For visualization purposes, in Figure~\ref{20130328_synthetic} we display the fit to a LASCO C2 image at an earlier time (20:24 UT), given that the CME appears too small in the LASCO C3 image close to 22:24 UT. To simulate the CME at 20:24 UT, basically the same parameters were used, but with $h_{front}$ = 6.7\,\Rsun. The event under study can be considered to expand in a self-similar manner, given that \citet{Subramanian-etal2014} interpreted variations of $\kappa$ as indicative of non-self-similar expansion. We prefer to abstain from further interpretations, however, since the simulation with the GCS model during the whole evolution of the event was not in the scope of this analysis.

The forward-modeling tool also allows to produce synthetic white-light images from the perspective of both STEREO spacecraft by assuming an electron distribution and Thomson scattering, as shown in the lower panel of Figure~\ref{20130328_synthetic}. The resemblance of the synthetic images to the observations is remarkable and confirms our direct interpretation from the observations, namely that COR2-B detects the axial view of the CME and COR2-A the lateral view.

\section{Expansion}\label{s:expansion}
\begin{figure}[!h]
  \centering
\includegraphics[width=1.\linewidth]{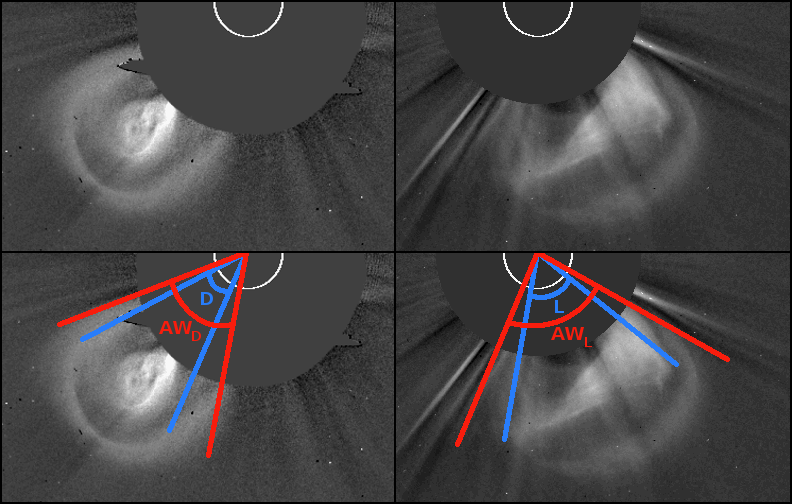}\llap{\parbox[b]{4.79in}{\colorbox{light-gray}{\color{white}{\tiny {\fontfamily{cmss}\selectfont COR2-B 20130328 20:54 UT}}}\\\rule{0ex}{2.891in}}}\llap{\parbox[b]{2.4in}{\colorbox{light-gray}{\color{white}{\tiny {\fontfamily{cmss}\selectfont COR2-A 20130328 20:54 UT}}}\\\rule{0ex}{2.891in}}}
\caption{Top: the CME exhibits its axial view in the perspective from ST-B (left) and the lateral view from ST-A (right). Bottom: same as above, but with colored lines indicating how the measurements of \textit{D} and \textit{L} (in blue) and \textit{AW$_D$} and \textit{AW$_L$} (in red) are performed.}
\label{aw}
\end{figure}

As stated in Section \ref{s:ID}, the orientation of the main axis of the CME on 28 March 2013 is highly favorable for observing its lateral perspective from ST-A and the axial one from ST-B. This allows the study of the evolution of the expansion in both directions, \ie along the flux rope's main axis and perpendicular to it. To perform this analysis, we followed two approaches. On the one hand, we studied the evolution of the magnitudes \textit{D} and \textit{L}, adopting the same criterion used in \citet{Cremades-Bothmer2005}, to allow for direct comparison with their results. \textit{D} represents the flux rope diameter by assuming the flux rope as an entity with cylindrical symmetry. It is measured as the angular distance of the inner edges of the cavity in the axial perspective, without considering the bright outer rim, as indicated by the blue lines in the bottom left panel of Figure~\ref{aw}. Likewise, \textit{L} refers to the length of the extended prominence material, which is generally accepted to be located at the bottom of the flux rope, aligned with the cylinder axis, and is measured as the angular extension of the inner bright and elongated feature, as  shown by the blue lines in the bottom right panel of Figure~\ref{aw}. Two snapshots of the CME axial and lateral views without the outlined angular extents are also shown in the top panel of Figure~\ref{aw} for visual comparison.  

On the other hand, the angular width (AW) of the external envelope of the CME from the lateral \textit{AW$_L$} and from the axial \textit{AW$_D$} perspectives was also measured, given that the angular width is a commonly measured attribute of CMEs. In the bottom panel of Figure~\ref{aw}, \textit{AW$_D$} and \textit{AW$_L$} are indicated by red lines to the left and right, respectively. Dimensions from both within and outside the envelope were considered and measured as angular distances with the vertex being located in the center of the solar disk. In practice, both \textit{D} and \textit{L} differ from \textit{AW$_D$} and \textit{AW$_L$}, respectively, because the outer rim of the CME has a significant width. The difference between \textit{D} and  \textit{AW$_D$} could be related to the amount of piled-up and compressed material in the lateral flanks of the CME as it expands. The measurements of \textit{D}, \textit{L}, \textit{AW$_D$}, and \textit{AW$_L$} were simultaneously performed to obtain the ratios \textit{L/D} and \textit{AW$_L$/AW$_D$} at every point in time.

\begin{figure}[!h]
  \centering
    \includegraphics[trim =60mm 0mm 20mm 100mm, clip,width=0.423\linewidth]{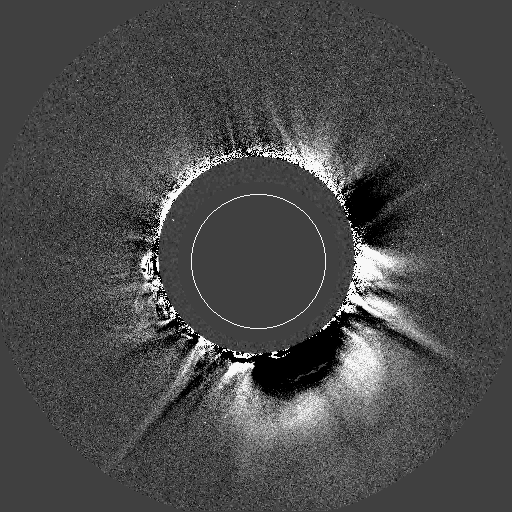}\llap{
  \parbox[b]{1.98in}{\tiny \cw{{\fontfamily{cmss}\selectfont COR1-A 20130328 18:55 UT}}\\\rule{0ex}{0.0in}}}
    \includegraphics[trim = 160mm 115mm 120mm 185mm, clip,width=0.45\linewidth]{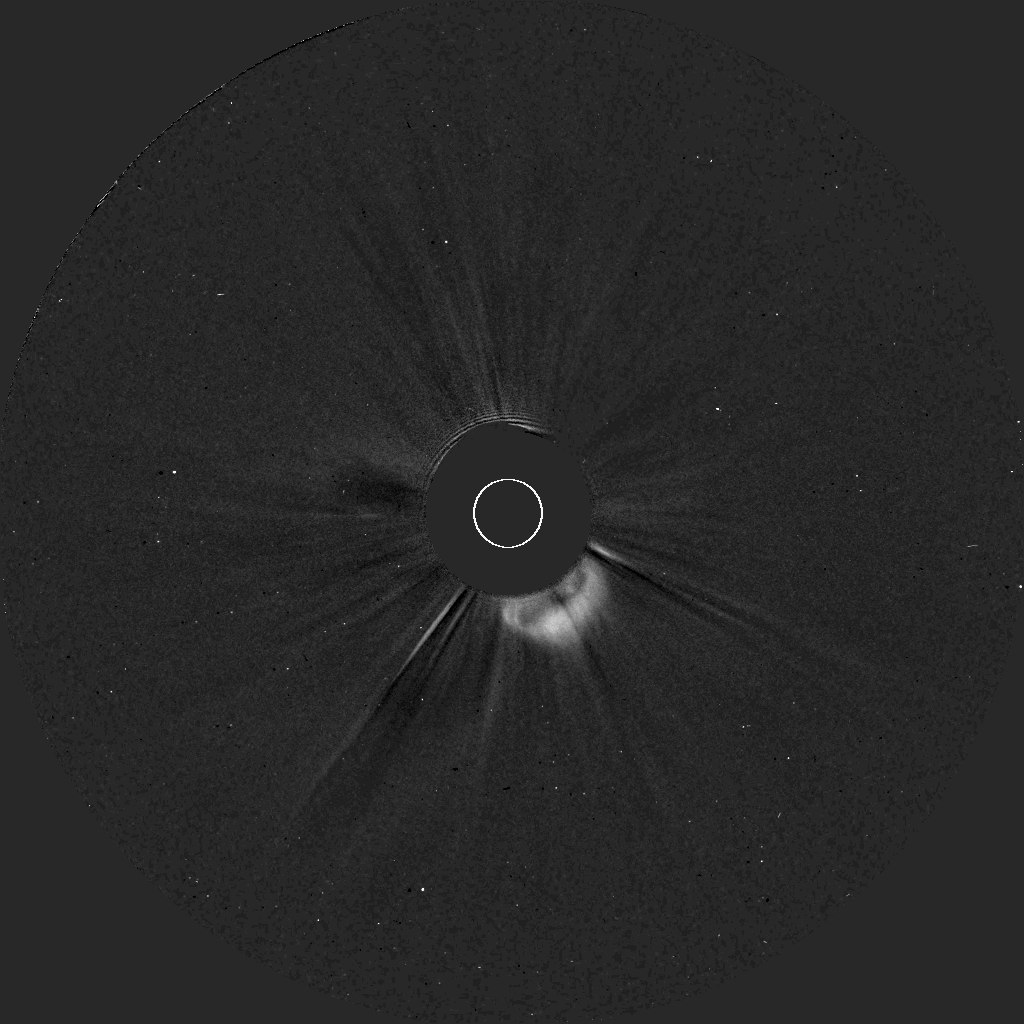}\llap{
  \parbox[b]{2.1in}{\tiny \cw{{\fontfamily{cmss}\selectfont COR2-A 20130328 18:54 UT}}\\\rule{0ex}{0.0in}}}
    \includegraphics[trim = 60mm 0mm 20mm 100mm, clip,width=0.423\linewidth]{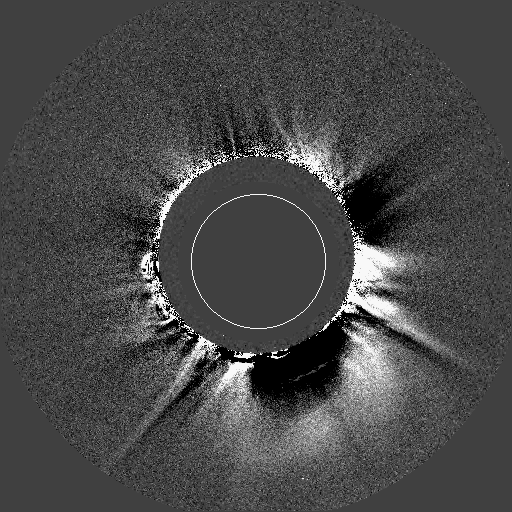}\llap{
  \parbox[b]{1.98in}{\tiny \cw{{\fontfamily{cmss}\selectfont COR1-A 20130328 19:25 UT}}\\\rule{0ex}{0.0in}}}
    \includegraphics[trim = 160mm 115mm 120mm 185mm, clip,width=0.45\linewidth]{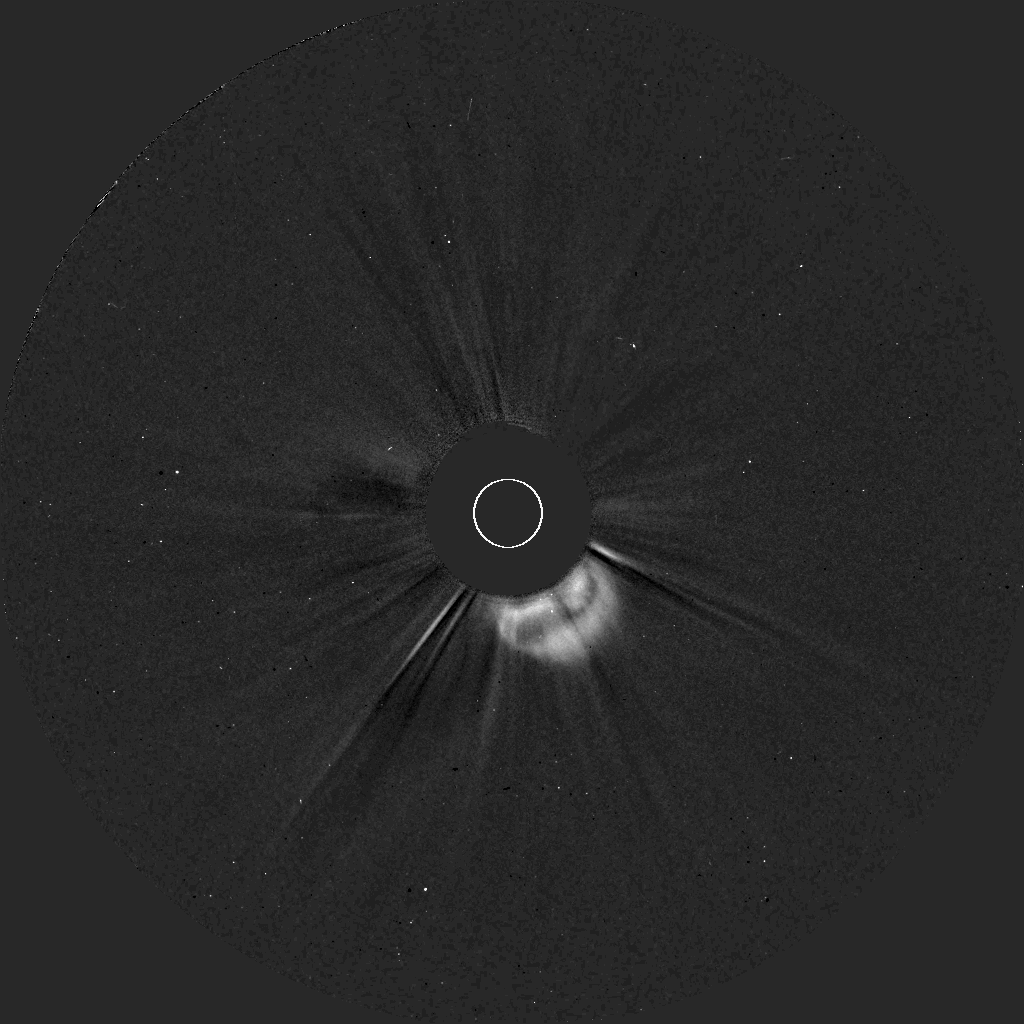}\llap{
  \parbox[b]{2.1in}{\tiny \cw{{\fontfamily{cmss}\selectfont COR2-A 20130328 19:24 UT}}\\\rule{0ex}{0.0pt}}}
    \includegraphics[trim = 60mm 0mm 20mm 100mm, clip,width=0.423\linewidth]{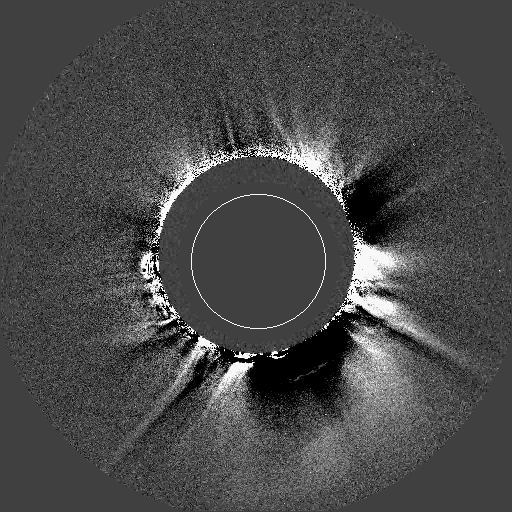}\llap{
  \parbox[b]{1.98in}{\tiny \cw{{\fontfamily{cmss}\selectfont COR1-A 20130328 19:55 UT}}\\\rule{0ex}{0.0in}}}
    \includegraphics[trim = 160mm 115mm 120mm 185mm, clip,width=0.45\linewidth]{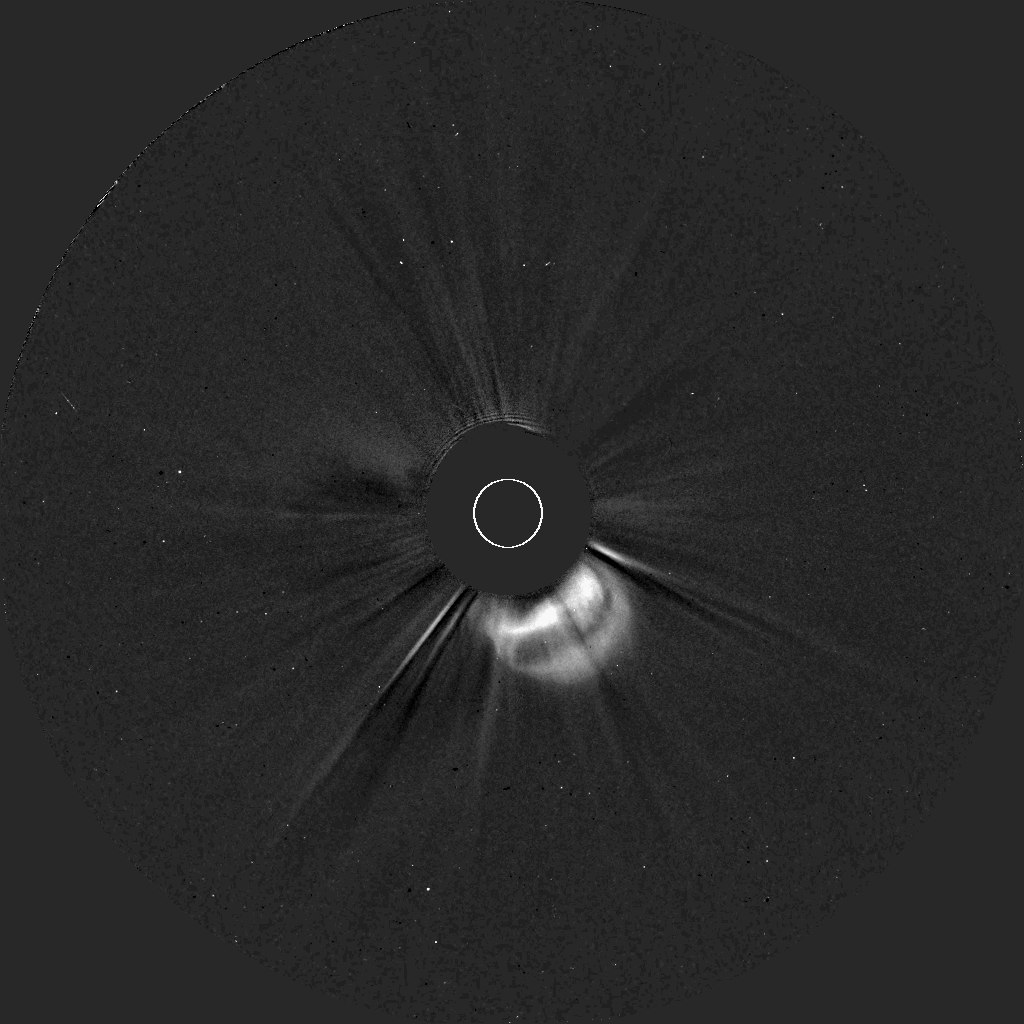}\llap{
  \parbox[b]{2.1in}{\tiny \cw{{\fontfamily{cmss}\selectfont COR2-A 20130328 19:54 UT}}\\\rule{0ex}{0.0in}}}
  
    \caption{The lateral perspective of the CME as seen in base-difference images from COR1-A (left) and COR2-A (right) at three different times: $\approx$\,18:55 UT, $\approx$\,19:25 UT, and $\approx$\,19:55 UT.}     
\label{flanks}
\end{figure}

Although the measurements of \textit{D}, \textit{L}, \textit{AW$_D$}, and \textit{AW$_L$} were obtained from difference images to highlight structures, great effort was made to ensure that they refer to the same structures in images from different instruments for each of the perspectives. Some difficulties were experienced because the same features were detected with different contrast in different instruments. An example of this can be seen in Figure~\ref{flanks}, where the flanks of the CME seen in COR1 are very faint and diffuse when compared to those observed in COR2 at approximately the same time. The same happens with the CME leading edge: it is almost imperceptible near the outer edge of the COR1 FOV at 18:55 UT, with the bright elongated feature corresponding to the prominence material. However, in a COR2 image at the same time, the leading edge is bright and well defined, while the bright prominence just begins to emerge from outside the occulter. At 19:25 and 19:55 UT, the leading edge is out of the COR1 FOV, and both prominence material and leading edge are well distinguishable in COR2. To overcome these obstacles, the sequence of images from each instrument was inspected back and forth several times, and compared with those from other instruments. During this process, we noted that the leading edge of the CME is not observed to emerge from the COR1 occulter, but rather to start its outward movement at an approximate height of 2\,\Rsun, leaving behind a dark feature in base-difference images (Figure~\ref{flanks}, left column), indicative of a structure that was located at that height before the eruption.

\begin{figure}[!h]
  \centering
  \includegraphics[trim = 14mm 10mm 2mm 14mm, clip,width=0.47\linewidth]{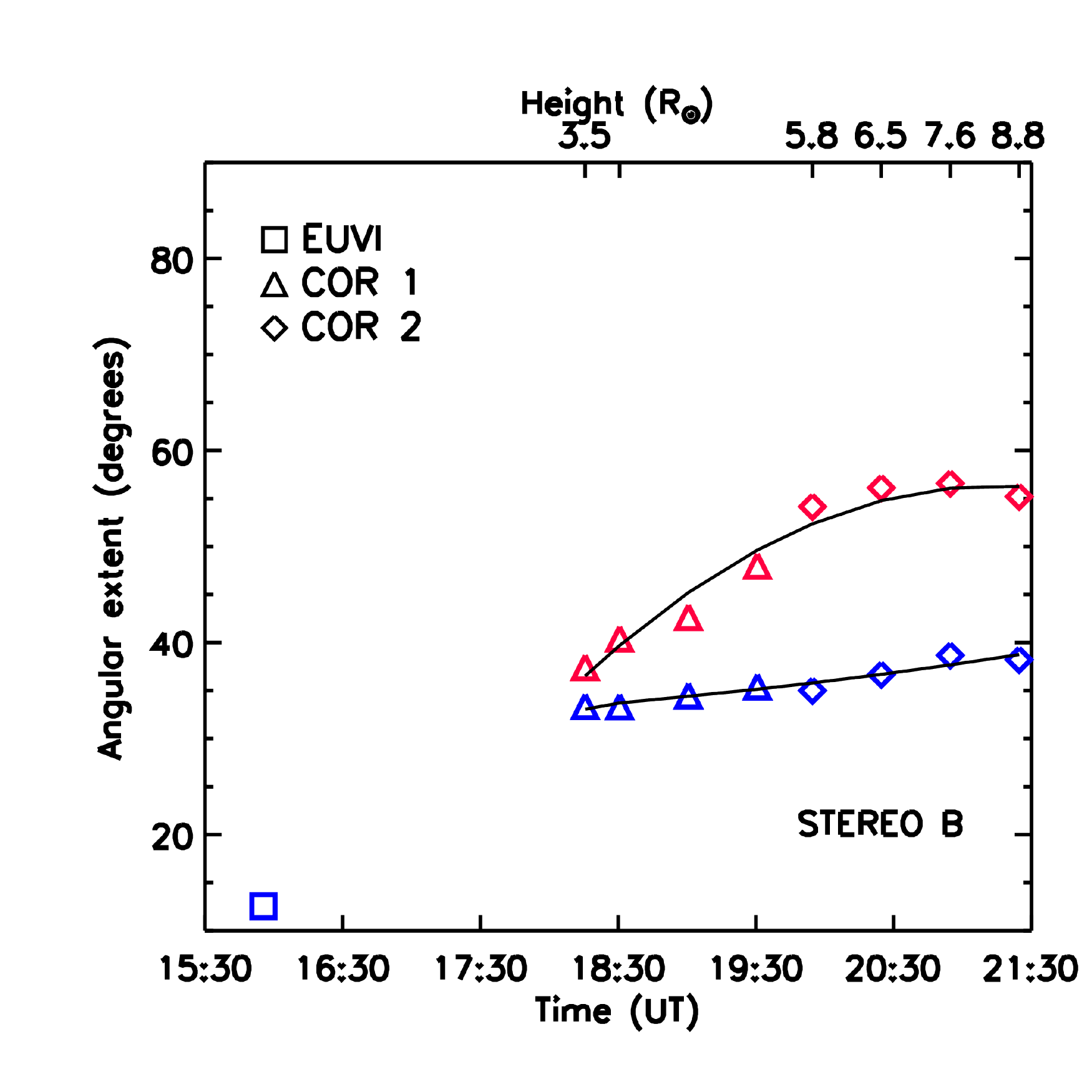}
  \includegraphics[trim = 14mm 10mm 2mm 14mm, clip,width=0.47\linewidth]{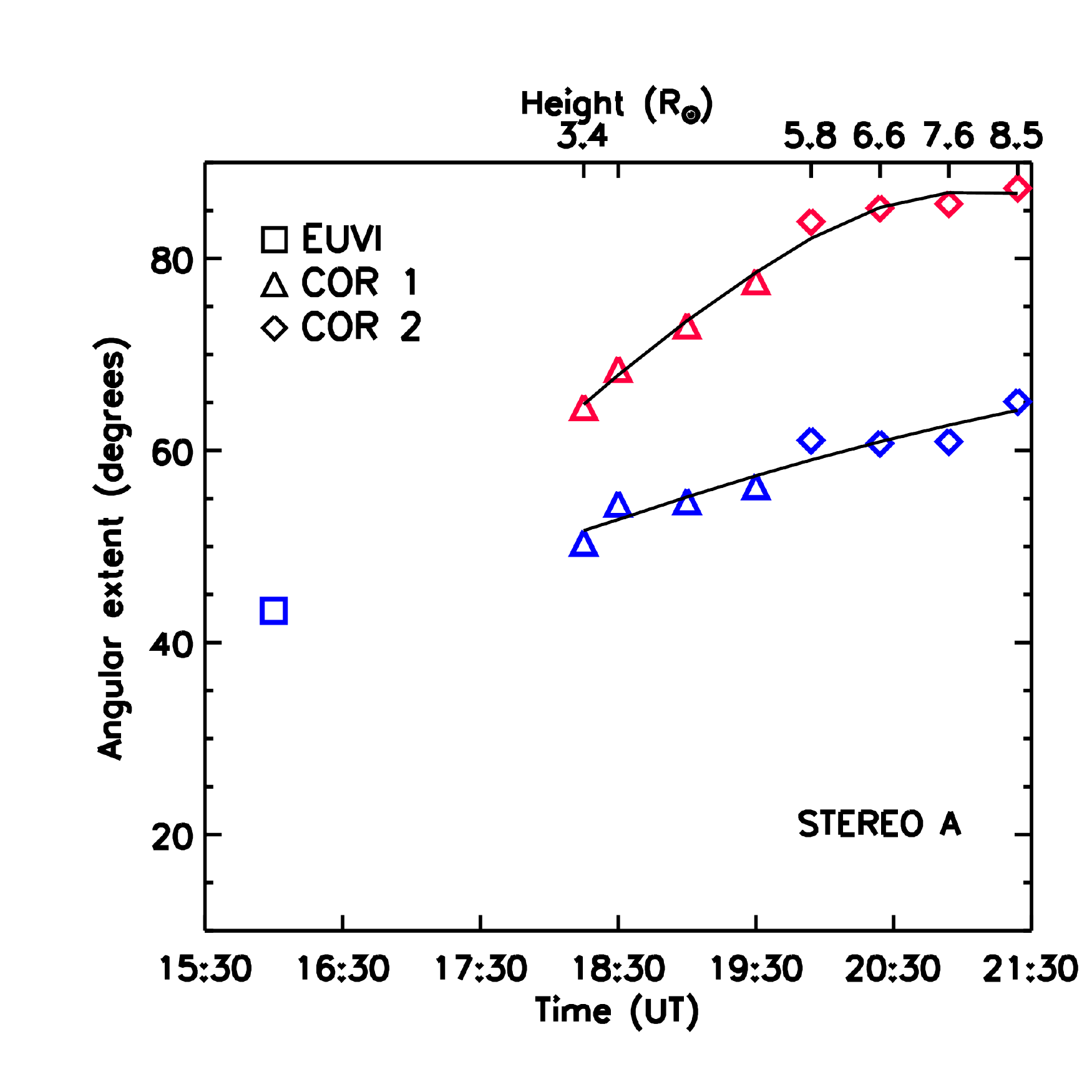}
\caption{Expansion of the studied CME as a function of time. Left: temporal evolution of \textit{D} (blue) and \textit{AW$_D$} (red) as measured in the axial view from ST-B observations. Right: temporal evolution of \textit{L} (blue) and \textit{AW$_L$} (red) as measured in the lateral view from ST-A. The solid black line represents the b-spline fit to the data to help visualize the general behavior discussed in the text. The top horizontal axis shows for reference the height of the CME leading edge.}
\label{width_height}
\end{figure}

The temporal evolution of the expansion of the CME in two directions, along and perpendicular to the main axis, is shown in Figure~\ref{width_height} as deduced from ST-B (left) and ST-A (right) data. The height of the leading edge of the CME is also indicated for each data point in the top horizontal axis. It is measured as the distance from the solar center to the outer end of the CME leading edge and expressed in units of solar radii ($R_{\odot}$).  It was not possible to measure the height of the low-coronal eruptive features because the outer end of the CME was not noticeable in the EUVI 195\,\AA~images. Width measurements from EUVI, COR1, and COR2 data are indicated with squares, triangles, and diamonds respectively. Blue and red denote the inner (\textit{D} and \textit{L}) and outer (\textit{AW$_D$} and \textit{AW$_L$}) angular extents. Since the selection of an appropriate model that fits the data points is a complex task, especially for the behavior of the outer angular extents, we used a piecewise polynomial function, also called ``b-spline", to fit the data so that the small fluctuations are smoothed out and the general trend is enhanced. This fit is shown in black in Figure~\ref{width_height}. For all the data sets we chose b-splines of order 3.

As shown in the left panel of Figure~\ref{width_height}, the \textit{AW$_D$} (red) related to the axial perspective observed in ST-B images grows linearly with time in the COR1-B FOV until $\approx$\,5$R_{\odot}$, keeping a nearly constant value afterwards. The inner angular extent \textit{D} (blue), on the other hand, shows a somewhat linear trend in the coronagraphs FOV. Only the angular extent \textit{D} could be measured in EUVI because the erupting structure seen in the low corona (see left panel in Figure~\ref{low_corona}) only shows the circular features believed to outline the flux rope. The large difference between the value of \textit{D} measured from EUVI observations and that from COR1 suggests an enormous expansion in the early stages of the CME, unfortunately not captured by EUVI due to a data gap from 16:00 to 18:00 UT.

Similarly, for the lateral perspective detected by ST-A in the right panel of Figure~\ref{width_height}, the inner and outer angular extents \textit{L} and \textit{AW$_L$} also show different behavior. The external values of \textit{AW$_L$} show a linear temporal evolution up to $\approx$\,5$R_{\odot}$ followed by a change in slope. As for the previous case, only the angular extent of \textit{L} could be measured in EUVI images because only the erupting prominence could be observed and not the flanks of the CME (Figure~\ref{low_corona}, right panel). The inner measure of \textit{L} grows linearly with time until the edge of the COR2-A FOV. The first values of \textit{L} measured in COR1-A images differ by almost 15$\degree$ from those of \textit{AW$_L$} at the same times, in contrast with a small difference of $\approx$5$\degree$ between the first measurements of \textit{D} and \textit{AW$_D$} in COR1-B. From Figure~\ref{width_height} it is evident that the extension in the lateral perspective and the expansion rate are larger than those in the axial perspective. The measured \textit{AW$_L$} and \textit{AW$_D$} values corresponding to 20:24 UT are 85$\,^{\circ}$ and 56$\,^{\circ}$, respectively. From the GCS parameters at the same time we can determine the AW corresponding to the lateral extent as \textit{AW$_{L(GCS)}$} = 2$\,(\alpha\,+\,\delta)$, which yields 88$\,^{\circ}$, and that of the axial extent as \textit{AW$_{D(GCS)}$} = 2$\,\delta$, resulting in 56$\,^{\circ}$. These values are very similar to those measured directly on the images, which is expected given that the main axis of this particular event is approximately parallel to the POS in the ST-A view, and likewise, almost perpendicular to the POS in the ST-B view.

The observed profile of the minor radius of the flux rope \vs time (blue symbols in Figure~\ref{width_height}, left panel) resembles the general behavior predicted by different theoretical approaches (see \eg, Figure 1 in \opencite{Chen-Garren1994}; Figure 8 in \opencite{Lin-etal2004}), showing a very rapid expansion during the first stages of the eruption followed by a deceleration phase as the CME leading edge reaches larger distances from the Sun. These approaches differ, however, in the dimensions and location of the assumed flux rope structure. On the other hand, the expansion along the main axis of symmetry is generally not discussed in detail in theoretical works, therefore we did not find similar examples that could be compared with our findings. We are confident that further studies like the one presented here will contribute to better constrain and therefore improve current models.

As previously addressed, these angular extents obtained from ST-B and ST-A represent the \emph{D} and \emph{L} parameters, respectively, of the three-dimensional configuration proposed by \cite{Cremades-Bothmer2004}. \cite{Cremades-Bothmer2005} averaged the values of \textit{D} and \textit{L} measured for several different events exhibiting only one of the perspectives. The ratio between the average value of \textit{D} deduced from one set of events, and the average value of \textit{L} deduced from other set of events, \ie the ratio of average lateral to axial dimensions, was found to be \emph{L/D} = 1.6. They also found a relationship between the length of the source region and the measure of \textit{L} of the corresponding CME, with \textit{L} being wider for large source regions, which tend to be located at higher latitudes. As stated above, the \emph{L/D} = 1.6 corresponds to the averages of separate measurements of \emph{L} and \emph{D} performed on different events. Here we report on the first simultaneous measurement of \emph{L} and \emph{D} for a CME event, and thus the first \emph{L/D} deduced for the same single CME. In addition, we also report the ratio \textit{AW$_L$}/\textit{AW$_D$} obtained from the full angular widths exhibited in the lateral and axial perspectives. Both the \emph{L/D} and \textit{AW$_L$}/\textit{AW$_D$} determined at several points in time yields $\approx\,1.6$ for the CME under study, the same value as deduced by \cite{Cremades-Bothmer2005}.

\section{Conclusions}\label{s:conclusions}

The hypothesis posed by \cite{Cremades-Bothmer2004}, according to which CMEs are organized along a main axis of symmetry and therefore should exhibit different appearances according to their location, orientation, and vantage point, is directly verified by the simultaneous observation of the two extreme perspectives relative to the same event. The analysis of the event was achieved by combining the stereoscopic views of STEREO and the terrestrial views of SOHO and SDO. With two spacecraft in quadrature, CMEs suitable to exhibit both perspectives are those that arise from polar regions and are directed perpendicular to the Sun--observer line. Such an event was identified in the images provided by the STEREO/SECCHI coronagraphs on 28 March 2013, with the STEREO spacecraft separated by $\approx$\,86$\degree$. The lateral and axial perspectives are unambiguously discerned in the fields of view of the ST-A and ST-B, respectively. The source region of this event could not be observed in detail from chromospheric or low-coronal images for several reasons: i) the source was on the far side for SDO/AIA and extreme limb for ST-A and ST-B, ii) the prominence associated with this CME was presumably suspended high in the low corona prior to eruption, which has been considered by \cite{Robbrecht-etal2009} to explain stealth CMEs, iii) if the latter is the case, the filament was probably too hot to be detected in H$\alpha$.

This event has a favorable orientation that allows for the direct detection of the lateral and axial perspectives and enables a temporal analysis of the CME expansion. The expansion of the flux rope angular diameter D measured in the axial perspective, as well as that of the lateral angular extent of the associated prominence, show a linear increase in time, at least up to the outer edge of the COR2 FOV. The full angular widths of the CME as seen in the axial (\textit{AW$_D$}) and lateral (\textit{AW$_L$}) perspectives show a different behavior in time: they show a linear growth with time up to $\approx$\,5$R_{\odot}$, followed by a slower growth rate phase in the case of the lateral perspective, and by a phase of nearly constant AW in the case of the axial view.

The average ratio $L/D$ obtained from values at different points in time yielded $\approx\,1.6$. This is the first time that this ratio is deduced for the same single CME, and it agrees with previous analyses obtained from measurements of single perspectives performed on different events. The average \textit{AW$_L$}/\textit{AW$_D$} of the full angular widths in the lateral and axial perspectives yields the same value.

A similar analysis performed on a set of nearly polar CMEs is underway. We hope to understand whether there are recurrent patterns regarding the distinct angular extents of the lateral and axial perspectives, as well as the expansion rates in the axial direction and perpendicular to it. As for equatorial CMEs, the simultaneous detection of both perspectives requires the combined analysis of coronagraphic observations offset from the ecliptic, such as those expected to be provided by the \textit{Solar Orbiter} mission, to be launched in October 2018, together with observations from close to the ecliptic plane, \eg~from Earth, SOHO, or \textit{Solar Probe Plus}, to be launched in July 2018.

\begin{acks}
IC acknowledges a postdoctoral fellowship from \mbox{CONICET}. HC and LB are members of the Carrera del Investigador Cient\'ifico (CONICET). The authors acknowledge funding from UTN project PID UTI2218 and thank the anonymous referee for valuable suggestions. The SOHO/LASCO data are produced by an international consortium of the NRL (USA), MPI f\"ur Sonnensystemforschung (Germany), Laboratoire d'Astronomie (France), and the University of Birmingham (UK). SOHO is a project of international cooperation between ESA and NASA. The STEREO/SECCHI project is an international consortium of the NRL, LMSAL and NASA/GSFC (USA), RAL and Univ. Bham (UK), MPS (Germany), CSL (Belgium), IOTA and IAS (France). SDO/AIA data are courtesy of the NASA/SDO and the AIA Science Teams. This article uses data from the SOHO/LASCO CME catalog generated and maintained at the CDAW Data Center by NASA and the CUA in cooperation with NRL.
\end{acks}

\noindent{\bf Disclosure of Potential Conflicts of Interest} The authors declare that they have no conflicts of interest.

\bibliographystyle{spr-mp-sola}
\bibliography{CCB}

\end{article}
\end{document}